\documentclass[a4paper,11pt]{article}
\usepackage{jheppub}

\usepackage[T1]{fontenc} 

\usepackage{setspace}
\usepackage{subcaption}
\usepackage{physics}
\usepackage{dsfont}
\usepackage{tensor}
\usepackage[normalem]{ulem}
\usepackage{xcolor}
\usepackage{graphicx}
\usepackage[export]{adjustbox}
\usepackage{mathtools}
\usepackage{mathbbol}
\usepackage{boondox-cal}
\usepackage[centertableaux]{ytableau}
\usepackage{sidenotes}
\usepackage{mathdots}
\usepackage{tikz}

\colorlet{darkgreen}{green!50!black}

\title{Entanglement Edge Modes of General Noncommutative Matrix Backgrounds}
   \author[a]{Alexander Frenkel}

   \affiliation[a] {Stanford Institute for Theoretical Physics, Stanford University,\\382 Via Pueblo, Stanford CA 94305}

   \emailAdd{afrenkel@stanford.edu}

   \abstract{
        We explore the structure of entanglement edge modes on noncommutative backgrounds that arise from matrix quantum mechanics. For the fuzzy sphere, despite nonlocality and UV/IR mixing, we find area law behavior in the dominant $U(N)$ representations governing the state of the edge modes. For general noncommutative backgrounds with no global symmetry, nonlocal effects resum into a smoothly varying coupling constant that deforms the metric to a different frame. The effect is analogous to the relationship between string frame and Einstein frame in string theory.
   }
\date{\today}

\begin{document}

\maketitle

\section{Introduction}\label{sec:intro}

Entanglement entropy has risen to a prominent role in the study of quantum gravity. At the heart of this phenomenon is the observation that the semiclassical path integral gives a UV-finite answer with area-law behavior, $A/4G_N$, for the entanglement between the interior and exterior of a black hole \cite{Srednicki:1993im, Susskind:1994sm}. A critical test for any proposed microscopic definition of quantum gravity is the reproduction of this result.

\begin{figure}[h]
\centering
\begin{subfigure}{.5\textwidth}
  \centering
  \includegraphics[width=0.9\linewidth]{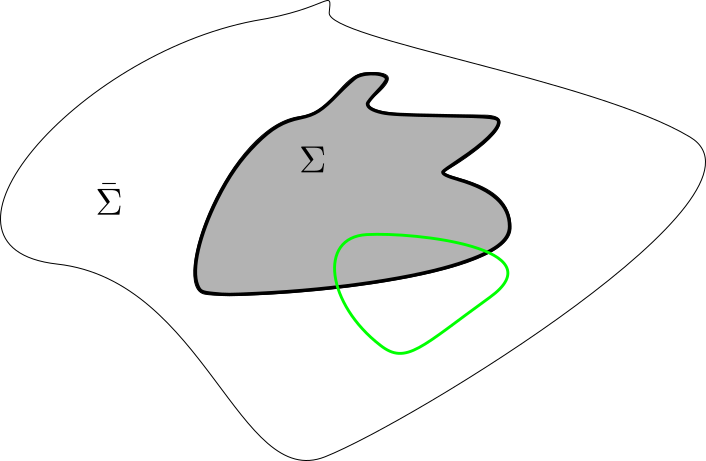}
  \caption{}
  \label{fig:custring1}
\end{subfigure}%
\begin{subfigure}{.5\textwidth}
  \centering
  \includegraphics[width=0.9\linewidth]{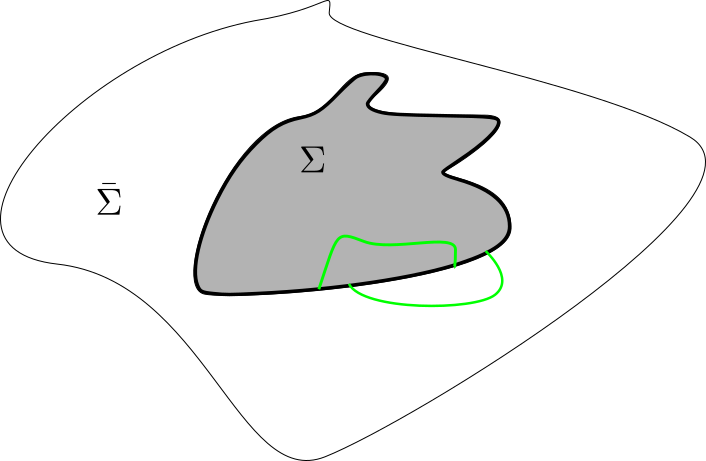}
  \caption{}
  \label{fig:cutstring2}
\end{subfigure}
\caption{On the left an extended degree of freedom (such as a Wilson loop in a gauge theory or a string in string theory) crosses an entanglement cut. The cut partitions the spatial geometry into subregion $\Sigma$ and its complement $\bar{\Sigma}$. On the right, we have extended the Hilbert space to include open Wilson lines or strings that are constrained to end on the entanglement cut.}
\label{fig:cutstrings}
\end{figure}

In 1993, Susskind and Uglum \cite{Susskind:1994sm} probed the question of entanglement in string theory. Among other results, they proposed two conjectures that continue to guide intuition in holography to this day: 
\begin{enumerate}
\item The divergent area-law entanglement of matter fields propagating in the fluctuating geometry of gravity renormalizes the bare Newton's coupling $G_{N0}$ to a finite renormalized value $G_{NR}$. The entanglement entropy in the full UV complete theory of quantum gravity is finite and its $O(1)$ coefficients are determined:
\begin{equation}
S_{EE} = \left(\frac{A}{4G_{N0}} + \frac{A}{\epsilon}\right) + O(\log A) = \frac{A}{4G_{NR}} + O(\log A).
\end{equation}

\item The Hilbert space of closed string theory does not factorize. Open strings anchored to the entangling surface must be introduced to consider the Hilbert space of a geometric subregion of the string target space (see Fig. \ref{fig:cutstrings}). 
\begin{equation}
\mathcal{H}_{closed} \subset \mathcal{H}_{open} \otimes \mathcal{H}_{open}.
\end{equation}

To restrict from $\mathcal{H}_{open} \otimes \mathcal{H}_{open}$ to $\mathcal{H}_{closed}$, we project onto configurations where the open strings on either side of the entanglement cut match up into closed strings in an appropriate manner.

\end{enumerate}

This second conjecture is intimately connected to the study of entanglement edge modes in gauge theories\footnote{See \cite{Donnelly:2020teo,Jiang:2020cqo} for an explicit instantiation in topological string theory.}. In gauge theories `large gauge transformations'\footnote{This term is really a misnomer, as these are global symmetries and not gauged.}, or gauge transformations that are supported on the boundary of the spacetime manifold the theory is defined on, are anomalous. They must therefore be promoted to physical degrees of freedom when quantizing the system \cite{Donnelly:2016auv}. When we introduce a boundary by inserting an entanglement cut, we introduce additional degrees of freedom on the entangling surface. In gauge theories, this means inserting boundary charges confined to propagate on the entangling surface - a classic example is a WZW boson propagating on the boundary of Chern-Simons gauge theory \cite{ELITZUR1989108,kitaev2006topological,balasubramanian2017multi}. See \cite{levin2006detecting,Soni:2015yga,Ghosh:2015iwa} for good reviews and explicit constructions of this phenomenon in lattice regularizations of gauge theories.

In this note, we focus on understanding string theory edge modes in the context of large-$N$ matrix quantum mechanics (MQM). Large-$N$ theories are relevant to string theory in two distinct but related ways. First, the `t Hooft organization of Feynman diagrams \cite{tHooft:1973alw} may be taken to define the topological expansion of the string partition function. Second, MQM appears as the effective worldline theory of $D0$-branes in type IIA string theory \cite{Elitzur:1997hc,Witten:1997ep}. 

In both cases, systems of interest have the structure of dimensionally reduced Super Yang-Mills (SYM) (see \cite{Taylor:2001vb} for a comprehensive review) -
\begin{equation}\label{eqn:MQM-lag}
L = \Tr[\sum_i \dot{X}^{i 2} + \sum_{i < j}[X^i,X^j]^2 + \ldots]. 
\end{equation}

The eigenvalues of $X^i$ carry the interpretation of positions of $D$-branes, and the off-diagonal elements represent strings propagating between the different branes. The commutator-squared interaction term is responsible for giving strings a potential energy linear in their length. The overall trace in \eqref{eqn:MQM-lag} makes manifest a global $SU(N)$ symmetry. This symmetry is typically taken to be gauged. Degrees of freedom that transform in the adjoint representation of $SU(N)$, such as the $X^i$, source Feynman diagrams with closed topology. For this reason $\mathcal{N=4}$ SYM \cite{Maldacena:1997re}, BFSS \cite{Banks:1996vh}, and BMN \cite{Berenstein:2002jq} are all dual to closed string theories, as they only contain adjoints. The mysterious string theory defined in the large-$N$ limit of QCD is contains open strings.

Equivalently, we may consider non-singlet states of the theory in which $U(N)$ is treated as a global symmetry, not a gauge symmetry. This choice is still holographically relevant \cite{Maldacena:2018vsr}. In this case, the non-singlet excitations will correspond to open string anchored to the boundary of the emergent geometry \cite{Marchesini:1979yq,Maldacena:2005hi,Maldacena:2018vsr}.

\begin{figure}[h]
\begin{center}
\includegraphics[width=0.5\textwidth]{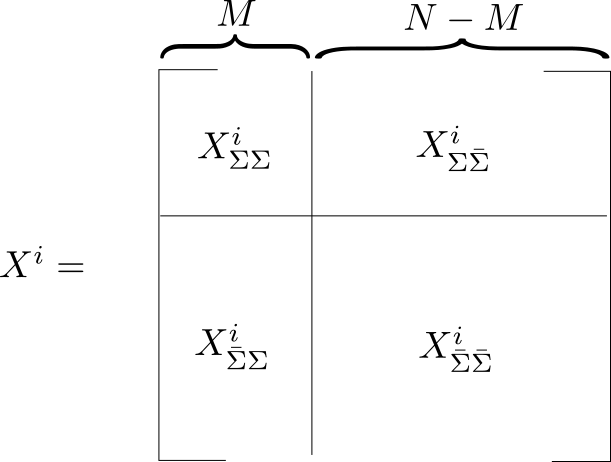}
\end{center}
\caption{The decomposition of the degrees of freedom we consider in this note. This decomposition is explained in detail in \S\ref{ssec:rev-ent}. In reference to Fig. \ref{fig:cutstrings}, $X^i_{\Sigma \Sigma}$ governs the internal degrees of freedom in subregion $\Sigma$, $X^i_{\bar{\Sigma}\bar{\Sigma}}$ the internal degrees of freedom of $\bar{\Sigma}$, and the mixed blocks $X^i_{\Sigma \bar{\Sigma}}$ and $X^i_{\bar{\Sigma}\Sigma}$ correspond to strings that cross the entanglement cut.} \label{fig:block-decomp}
\end{figure}

\begin{figure}[h]
\begin{center}
\includegraphics[width=0.7\textwidth]{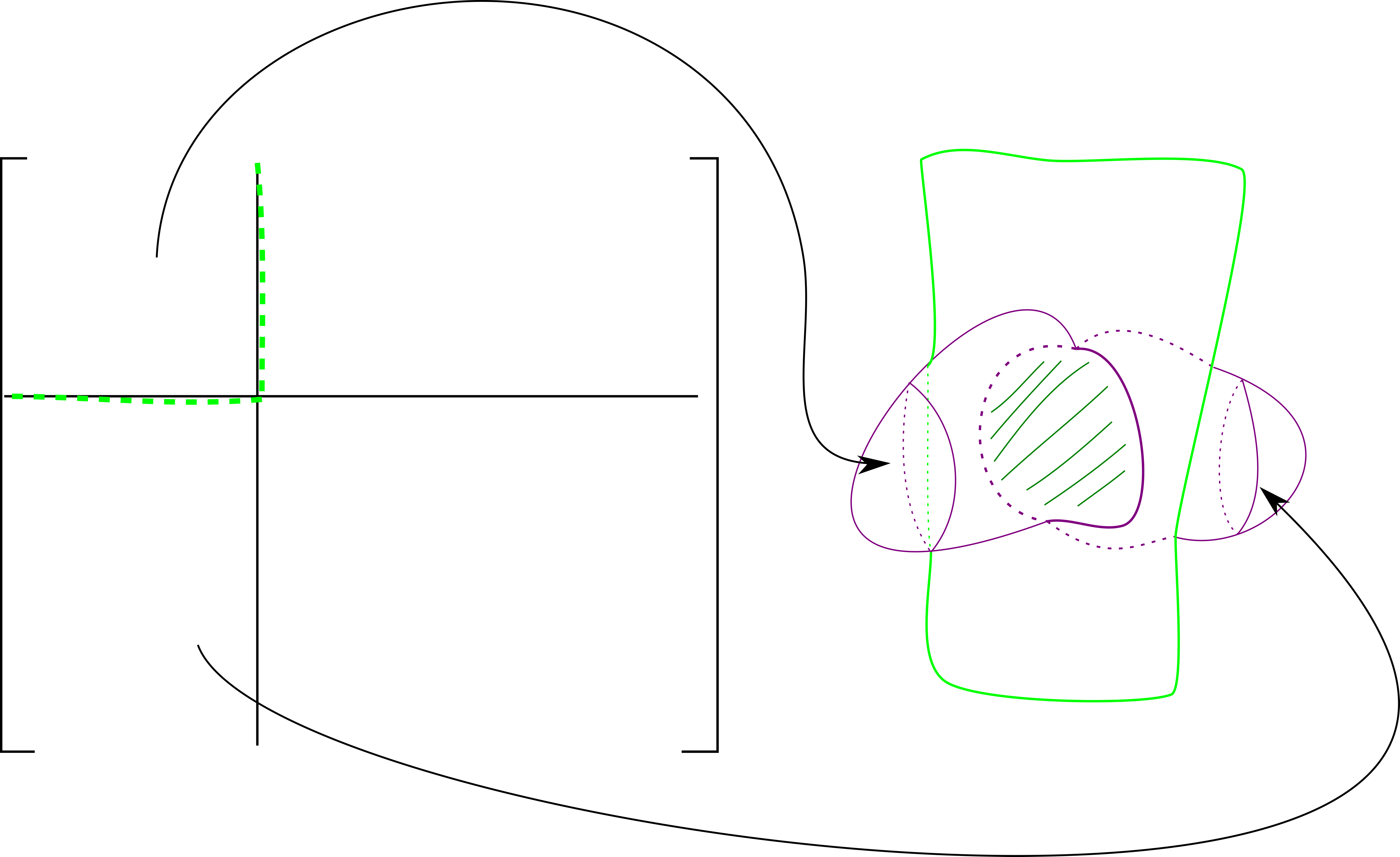}
\caption{On the left, we have decomposed the matrices $X^i$ into the $\Sigma \Sigma$ block and the rest of the system as in Fig. \ref{fig:block-decomp}. On the right is a cartoon of a sphere Feynman diagram cut open by an entangling surface. The $\Sigma \Sigma$ block is generically in a non-singlet state, so its internal correlation functions will generate open string Feynman diagrams. The arrows indicate that the open strings generated by $X^i_{\Sigma \Sigma}$ adjoints of the subregion and the $X^i_{\bar{\Sigma}\Sigma}$ $X^i_{\Sigma \bar{\Sigma}}$ bifundamentals match up to combine into closed string diagrams on the full space.}
\label{fig:open-string-gen}
\end{center}
\end{figure}

When we partition the matrix degrees of freedom into blocks (see Fig. \ref{fig:block-decomp}), the introduction of this partition breaks the global $U(N)$ symmetry down to $U(M) \cross U(N-M)$. This is precisely analogous to the introduction of an entanglement cut on the base space of gauged field theories. The off-diagonal blocks transform as bifundamentals under $U(M) \cross U(N-M)$, so will source open string endpoints to Feynman diagrams. Because the full theory contains only closed strings, we interpret the resulting open strings as being anchored to the entangling surface in the dual field theory (see Fig. \ref{fig:open-string-gen}). This is very similar in spirit to the cutting of closed string ribbon diagrams that happens in \cite{Horava:2020apz,Horava:2020she,Horava:2020val,Jafferis:2021ywg}.

In this note, we make this construction precise for MQM states that localize to noncommutative backgrounds. While not in the limit of fixed `t Hooft coupling relevant to holography, these states still of great interest in string theory and quantum gravity \cite{Szabo:2001kg,Douglas:2001ba,Steinacker_2010}. The worldvolume theory of $D$-branes with flux is often well descibed by noncommutative field theory \cite{ho1997noncommutative}. Physics on the compact $S^n$ of AdS$_m \cross S^n$ exhibits many features of noncommutative geometry \cite{douglas1998d,ho1997noncommutative} such as the stringy exclusion principle and giant gravitons \cite{McGreevy:2000cw}. Perhaps most surprisingly, the one-loop effective action for $3+1d$ noncommutative backgrounds in IKKT has recently been show to contain the Einstein-Hilbert term as an induced theory of gravity \cite{Steinacker:2023myp}. 

There has been much prior work on the entanglement in MQM and entanglement of noncommutative spaces (and the fuzzy sphere in particular) in the past. \cite{Das:1995vj,Das:1995jw} (and more recently \cite{Hartnoll:2015fca,Das:2022nxo}) consider entanglement in the $c=1$ matrix model, where in particular they find the entanglement entropy is regularized by precisely the emergent target space coupling constant. \cite{Karczmarek:2013jca,Han:2019wue} consider entanglement entropy on the fuzzy sphere. These two works consider the entanglement entropy of a scalar field on the surface of the sphere, so lose the gauge theoretic structure that gives rise to the edge modes we study. We interpret these works as the bulk corrections to the edge mode contribution we calculate. \cite{Karczmarek:2013jca} in particular find a transition from an area law to volume law behavior as they tune a UV regulator - we find similar behavior for the edge mode entanglement. \cite{Mazenc:2019ety,Das:2020jhy,Das:2020xoa,Hampapura:2020hfg} define target space entanglement entropy entanglement entropy and argue for its relevance in holographic systems. \cite{Hampapura:2020hfg} in particular draws a tantalizing connection to lattice gauge theories. \cite{Gautam:2022akq} defines a distinct but related choice of subsystems in MQM. \cite{Frenkel:2021yql,Frenkel:2023aft} consider entanglement edge modes in the particular cases of matrix quantum hall systems and mini-BMN.

\subsection{Layout of Note, Statement of Results, and Outline of Logic}

We now introduce each section, and along the way give a brief overview of the results and logic contained in each.

In \textbf{section \ref{ssec:rev-ncb}} we review the construction of noncommutative manifolds and entanglement edge modes, focusing on the context of MQM. In particular, we review how a set of classical $N \times N$ Hermitian matrices $X^i_{cl}$, $i \in \{1 \ldots D\}$, generate the algebra of noncommutative functions of a noncommutative manifold $\mathcal{M}_{\theta}$. These matrices may not all be independent, and may satisfy constraints of the form $f_a(X^i)=0$ (for example, the three matrices generating the fuzzy sphere algebra satisfy $\sum_i {X_{cl}^i}^2=0$). If we have $r$ constraints, the dimension of $\mathcal{M}_{\theta}$ will be $d = D-r$. The matrices $X^i_{cl}$ represent the coordinate functions on the manifold. Functions on $\mathcal{M}_{\theta}$ are given by polynomials in the $X^i_{cl}$, so we have a one to one correspondence between $N \times N$ matrices $F$ and functions on $\mathcal{M}_{\theta}$, which we denote $f(\vec{x})$. We write this correspondence as
\begin{equation}
f(\vec{x}) \leftrightarrow F.
\end{equation}

Derivatives on $\mathcal{M}_{\theta}$ are given by commutators with the coordinate functions $[X^i,F]$. In particular, the noncommutative laplacian is
\begin{equation}
\Delta f(\vec{x}) \leftrightarrow \frac{1}{N^2}\sum_{i=1}^D[X^i_{cl}[X^i_{cl},F]].
\end{equation}
Integrals are given by traces:
\begin{equation}
\int_{\mathcal{M}_{\theta}} \sqrt{G}d^d\sigma\, f(\vec{\sigma}) \leftrightarrow \frac{1}{N}\Tr[F].
\end{equation}
We have introduced $d$ independent coordinates $\sigma^a$ that parameterize the surface of $\mathcal{M}_{\theta}$. We will always be working in the matrix representations of noncommutative functions, so we leave the $\sigma^a$ implicit.

In \textbf{section \ref{ssec:rev-ent}}, we review entanglement edge modes specifically in the context of matrix quantum mechanics as presented in \cite{Frenkel:2021yql,Frenkel:2023aft}. Geometric subregions of gauge theories generically have reduced density matrices that are supported on nontrivial irreps of the gauge symmetry. This is true even when the global theory is a pure gauge theory with no matter degrees of freedom and is itself in a singlet state. The most direct way to see that this must be true, and the approach we take in this note, is to write the operator that measures the charge content of a subregion $\Sigma$ (which we call $Q_{\Sigma}$), and evaluate the expectation value $\langle Q_{\Sigma}^2 \rangle$ in the global pure state. In the commutative gauge theories we are accustomed to, this expectation value is proportional to the boundary area of the subregion, $|\partial \Sigma|$ (again see the discussions in \cite{Soni:2015yga,Ghosh:2015iwa,Donnelly:2016auv}). For the noncommutative gauge theories we study, we find a subtler relationship between the subregion's charge and its geometric data.

In \textbf{section \ref{ssec:u1}} we specialize to an arbitrary subregion on the fuzzy sphere, extending the results of \cite{Frenkel:2023aft}. Despite nonlocal effects due to UV/IR mixing, we demonstrate area-law behavior of the dominant boundary edge mode representations for arbitrary subregions. We focus on states that are localized to some classical state in the sense that the wavefunction $\psi(X^i)$ is a quadratic expansion around some classical minimum as 
\begin{equation}
X^i = X^i_{cl} + \sum_a \omega_a \delta x_a Y^i_a \implies K_{ij}, \quad \psi(\delta x_a) = \mathcal{N}\prod_a e^{-\omega_a \delta x_a^2}.
\end{equation}

The main result is that for any subregion of any topology (e.g. as pictured in Fig. \ref{fig:multicut}), the dominant irrep in the $SU(M)$ entanglement edge mode density matrix is given by a Young diagram of the form shown in Fig. \ref{fig:fuzz-sphere-yt}. The Young diagram has one row for each disconnected component of the entanglement cut, and the length of each row is given by the boundary length (i.e. `area') of this component. We show this by demonstrating that the second Casimir of the edge modes, $\langle \Tr[Q_{\Sigma}^2] \rangle$, is given by a matrix element of a second order differential operator on the fuzzy sphere. We express this operator in terms of a $3 \times 3$ set of fuzzy sphere functions $K_{ij}$ :
\begin{equation}\label{eqn:sec-cas-1}
\langle \Tr[Q_{\Sigma}^2] \rangle = \alpha N\Tr[\Theta_{\Sigma}K_{ij}[X^i,[X^j,\Theta_{\Sigma}]]] \leftrightarrow \alpha N^4 \int_{S^2}\sqrt{G}d^2\sigma \chi_{\Sigma} \kappa_{ij}(\vec{\sigma})L_iL_j \chi_{\Sigma}.
\end{equation}

Recall that $\Theta_{\Sigma}$ is the projector to the subregion, $\chi_{\Sigma}$ is the corresponding commutative characteristic function, and $L_i$ is the angular momentum operator on the fuzzy sphere. $\alpha$ is some $O(1)$ constant we calculate, and $\chi_{\Sigma}$ is the characteristic function of the subregion $\Sigma$. The functions $K_{ij}$ are calculated in terms of the conjugate momenta of the $X^i$ as
\begin{equation}\label{eqn:kij-def-1}
K_{ij} = \langle \Pi^i \Pi^j \rangle \leftrightarrow \kappa_{ij}(\vec{\sigma}).
\end{equation}
As $\chi_{\Sigma}$ is a step function, $L_iL_j \chi_{\Sigma}$ is a derivative of a delta function. Properties of noncommutative delta functions along with $SO(3)$ invariance ensure that the only role of the $\kappa_{ij}$ is to rescale $\alpha$ to $\tilde{\alpha}$:
\begin{equation}
\left \langle \Tr[Q_{\Sigma}^{2p}] \right \rangle = \tilde{\alpha}^p N^{2p} |\partial\Sigma|^{2p}.
\end{equation}
This immediately implies $Q_{\Sigma}$ has one eigenvalue for each disconnected component of $\partial \Sigma$, proportional in magnitude to the length of this disconnected component. We have therefore obtained the boundary entanglement for any region.

The states considered above describe emergent $U(1)$ noncommutative gauge theory. In \textbf{section \ref{ssec:uq}} we extend this calculation to backgrounds of emergent $U(q)$ gauge theories on the fuzzy sphere. The dominant Young diagram will now have $q$ rows for every disconnected component of $\partial \Sigma$, as pictured in Fig. \ref{fig:fuzz-sphere-yt-2}.

In \textbf{section \ref{sec:gen-mani}} we demonstrate how our results extend to arbitrary noncommutative manifolds $\mathcal{M}_{\theta}$, such as that pictured in Fig. \ref{fig:gen-mani}. We consider $\mathcal{M}_{\theta}$ whose curvature lengthscale. To obtain a similar structure to the fuzzy sphere we need the following two assumptions: the quantum state is givin by quadratic fluctuations around the classical geometry, and the corresponding free theory has at most second order kinetic terms.

The second Casimir may again be written in form \eqref{eqn:sec-cas-1} and \eqref{eqn:kij-def-1}. The functions $K_{ij}$ are dominated by the high energy normal modes around around the classical $\mathcal{M}_{\theta}$. Due to the UV/IR mixing inherent to noncommutative geometries will in general be a nonlocal function of the metric $G$. Despite this effect, we argue that (just like in the case of the fuzzy sphere) the corresponding $\kappa_{ij}(\vec{\sigma})$ are slowly varying in the sense that their derivatives are on the order of the curvature scale. We make use of eigenstate thermalization (more specifically a noncommutative analog of Berry's conjecture) \cite{srednicki1994chaos,berry1977regular,berry1983chaotic} to argue that the sum over UV normal modes may be treated as a random walk and average out to functions that vary on the curvature lengthscale of $\mathcal{M}_{\theta}$. 

In the case of the fuzzy sphere, we may absorb the $\kappa_{ij}$ into $\tilde{\alpha}$. For general $\mathcal{M}_{\theta}$ with no global symmetry, the deformation of the metric is more complex. Because \eqref{eqn:sec-cas-1} is a second order differential operator, we may define a new metric $\tilde{G}$ on $\mathcal{M}_{\theta}$ that satisfies the set of differential equations
\begin{equation}
\sum_{ab}\tilde{G}^{ab}\tilde{\nabla}_a \tilde{\nabla}_b f(\vec{\sigma}) = \sum_{ab}\kappa_{ab}G^{ab} \nabla_a \nabla_b f(\vec{\sigma}) \quad \forall\, f(\vec{\sigma}).
\end{equation}
$\tilde{G}_{ab}$ is locally related to $G_{ab}$ via some orthogonal matrix $O$ as $\tilde{G} = OGO^{\dag}$. In this sense, it may be interpreted as a change of frame. The dominant edge mode irrep will in general have area law behavior with respect to $\tilde{G}$, but not necessarily $G$. We draw the analogy to string theory, where propagating fundamental strings see the string frame metric but black hole entropy is an area law with respect to the Einstein frame metric.
 
In \textbf{section \ref{sec:discussion}} we discuss the results and comment on further work.

\section{Review}\label{sec:review}

\subsection{Noncommutative Backgrounds in MQM}\label{ssec:rev-ncb}

Noncommutative differential manifolds, much like their commutative counterparts, are defined via their algebra of functions, metric, and differential structure. The key distinction is that the multiplication operation on the algebra of functions is not assumed to be abelian. In this section, we review the construction of noncommutative geometries and their role in MQM as presented in e.g. \cite{Connes:1994yd,Connes:1998ngmt,Douglas:2001ba,steinacker2011non}. 

The nonabelian multiplication operation is often denoted as $\star$. We will use $\mathcal{M},\mathcal{N},\ldots$ to refer to commutative manifolds and $\mathcal{M}_{\theta},\mathcal{N}_{\theta}, \ldots$ for their noncommutative analogs. For example, the fuzzy plane is denoted by $\mathbb{R}^2_{\theta}$ and the fuzzy sphere by $S^2_{\theta}$. The corresponding algebras of functions will be denoted by $\mathfrak{a}_{\mathcal{M}}$ or $\mathfrak{a}_{\mathcal{M}_{\theta}}$. For the cases we are interested in these will be algebras over $\mathbb{C}$. The prototypical example of a noncommutative structure is the Moyal product defining the fuzzy plane \cite{moyal1949quantum}. For this historical reason we often refer to the product structure as the Moyal product. Similarly, when a matrix $F$ represents a particular noncommutative function we refer to this association as the Moyal map and denote it with a $\leftrightarrow$ as
\begin{equation}
F \leftrightarrow f(\vec{x}).
\end{equation}

The relevant commutative analogs to keep in mind are analytic manifolds, where the functions form an algebra over $\mathbb{C}$ that is generated by all sums and products of some the coordinate functions $\sigma^a$. For a $d$-dimensional manifold, $a$ runs from $1$ to $d$. We will repeat the same construction for noncommutative manifolds - we choose coordinate functions $\sigma^a$ that obey some commutation relations
\begin{equation}\label{eqn:ncm-alg}
\sigma^a \star \sigma^b - \sigma^b \star \sigma^a = [\sigma^a,\sigma^b] =: \theta^{ab}(\vec{\sigma}), \quad \theta(\vec{\sigma}):=\text{det}{}'\theta^{ab}(\vec{\sigma}).
\end{equation}
$\theta^{ab}(\vec{\sigma})$ is some antisymmetric two-form and det${}'$ is the pseudo-determinant (the product of all nonzero singular values). We refer to $\theta(\vec{\sigma})$ as the noncommutativity parameter. As reviewed in detail in \cite{steinacker2011non}, noncommutative manifolds defined this way are intimately related to the geometric quantization of symplectic manifolds. $\mathfrak{a}_{\mathcal{M}_{\theta}}$ is now defined as the quotient of the free algebra generated by all possible sums of all possible strings of $\sigma^a$ modulo the commutation relations in \eqref{eqn:ncm-alg}. We denote element of $\mathfrak{a}_{\mathcal{M}_{\theta}}$ by $f(\vec{\sigma})$. 

Noncommutative manifolds come equipped with a natural differential structure. Precisely because commutators obey the Leibniz rule, we may use \eqref{eqn:ncm-alg} to define
\begin{equation}
\begin{split}
&[\sigma^a, \sigma^b] = \theta^{ab}(\vec{\sigma}) \implies \theta^{ab}\partial_b f(\vec{\sigma}) := [\sigma^a,f(\vec{\sigma})].\\
&[\sigma^a,f(\vec{\sigma})g(\vec{\sigma})] = (\theta^{ab}\partial_b f(\vec{\sigma}))g(\vec{\sigma}) + f(\vec{\sigma})\theta^{ab}\partial_b g(\vec{\sigma}).
\end{split}
\end{equation}
In particular, this definition has the nice property
\begin{equation}\label{eqn:nc-deriv}
\theta^{ab}(\vec{\sigma})\partial_b \sigma^c = \theta^{ac}(\vec{\sigma}).
\end{equation}
The example of the fuzzy plane $\mathbb{R}^2_{\theta}$ is particularly illuminating in this context, as we have
\begin{equation}\label{eqn:nc-int}
[x,y] = \theta \implies [x,\cdot] \leftrightarrow \theta \partial_y \text{ and } [y,\cdot] \leftrightarrow - \theta\partial_x. 
\end{equation}

Finally, for the cases we are interested in $\mathfrak{a}_{\mathcal{M}_{\theta}}$ will be type I von-Neumann algebras (see \cite{Witten:2018zxz,Sorce:2023fdx} for recent reviews). As such, they come equipped with a trace, which we take to be normalized such that $\Tr[1] = 1$. The linearity of the trace makes it a natural candidate to define integration on $\mathcal{M}_{\theta}$, in the sense that
\begin{equation}
\int \mu(\vec{\sigma}) d^d\sigma\, f(\vec{\sigma}):= \Tr[f(\vec{\sigma})].
\end{equation}
We have introduced a symbol $\mu(\vec{\sigma})$ to denote the integration measure on $\mathcal{M}_{\theta}$. A priori the function $\mu(\vec{\sigma})$ is not independently defined and may just be thought of as notational dressing, but often tends to a nice functional form in the commutative limit ($\theta \rightarrow 0$) which may be deduced from inserting convenient test functions into the trace. The definition \eqref{eqn:nc-int} interacts nicely with our definition of the noncommutative derivative in \eqref{eqn:nc-deriv}. The cyclic property of the trace implies
\begin{equation}
\Tr[[f_1(\vec{\sigma}),f_2(\vec{\sigma})]f_3(\vec{\sigma})] = -\Tr[f_2(\vec{\sigma})[f_1(\vec{\sigma}),f_3(\vec{\sigma})]] \quad \forall f_1,f_2,f_3 \in \mathfrak{a}_{\mathcal{M}_{\theta}}.
\end{equation}
Remembering that commutators should be thought of as derivatives, this is simply integration by parts.

To make contact with MQM, we are interested in noncommutative manifolds whose algebra of functions may be represented by $GL(N,\mathbb{C})$\footnote{In some cases the algebra may be represented by a subgroup of $GL(N,\mathbb{C})$ but in most simple examples the entire group is generated. For our purposes, it is unimportant if we are working with all of $GL(N,\mathbb{C})$ or a subgroup.}. Generally, these are compact noncommutative manifolds with no boundary\footnote{Boundaries may be treated rather straighforwardly by introducing vectors in addition to the matrices, see e.g. \cite{Frenkel:2021yql,tong2015quantum,dorey2016matrix}.}. A representation of $\mathfrak{a}_{\mathcal{M}_{\theta}}$ is a homomorphism $r:\mathfrak{a}_{\mathcal{M}_{\theta}} \rightarrow GL(N,\mathbb{C})$ that preserves the trace and noncommutative structure in the sense that
\begin{equation}
\begin{split}
r([f_1(\vec{\sigma}),f_2(\vec{\sigma})]) &= [r(f_1(\vec{\sigma})),r(f_2(\vec{\sigma}))] \quad \forall f_1(\vec{\sigma}),f_2(\vec{\sigma}) \in \mathfrak{a}_{\mathcal{M}_{\theta}},\\
\Tr[r(f(\vec{\sigma}))] &= \Tr[f(\vec{\sigma})] \quad \forall f(\vec{\sigma}) \in \mathfrak{a}_{\mathcal{M}_{\theta}}.
\end{split}
\end{equation}
To ensure the second line is true, we must normalize the trace in $GL(N,\mathbb{C})$ by a factor of $1/N$.

In practice, we start from a representation and let it implicitly define a noncommutative manifold. We do this by choosing a collection of $D \geq d$ matrices $X^i \in GL(N,\mathbb{C})$. These matrices will satisfy some commutation relation and constraint equations
\begin{equation}
[X^i,X^j] = \tilde{\theta}^{ij}(\vec{X}), \quad h^r(\vec{X}) = 0. 
\end{equation}
We have introduced two new indices $i \in \{1,\ldots,D\}$ and $r \in \{1,\ldots,D-d\}$. To leading order in $\theta$, the relationship between $\tilde{\theta}^{ij}$ and $\theta^{ab}(\vec{\sigma})$ may be deduced from the \eqref{eqn:nc-deriv} and the chain rule:
\begin{equation}
\tilde{\theta}^{ij} = \theta^{ab}\partial_aX^i\partial_bX^j.
\end{equation}
This expression should not be taken too literally, but provides intuition about the relationship between $X^i$ and $\sigma^a$ in the commutative limit. We obtain useful intuition from considering some coordinate (say $X^1$) in its diagonal basis with eigenvalues ordered. We call the eigenvalues $x^1_n$, $n \in \{1,\ldots, N\}$. Smooth geometries tend to emerge when the eigenvalue distribution $\rho(x^1)$ is smooth in the large-$N$ limit. For functions of one variable, we directly have
\begin{equation}
\Tr[f(X^1)] = \frac{1}{N}\sum_{n=1}^N f(x^1_n) = \frac{1}{N}\int_{-\infty}^{\infty} \rho(x^1) dx^1\,f(x^1).
\end{equation}

Our interest is in manifolds where the noncommutativity parameter is small, so we may work perturbatively in $\theta^{ab}$. In the context of matrix representations, this means if we consider $X^1$ in its diagonal basis the remaining coordinates $X^i$ will be approximately supported only on diagonals near the main diagonal. It may be checked by explicit matrix multiplication that this is enough to ensure
\begin{equation}
\Tr[f_1(X^i)f_2(X^i)f_3(X^i)] = \Tr[f_2(X^i)f_1(X^i)f_3(X^i)] + O(p\theta),
\end{equation}
where $p$ is the largest power appearing in the monomial expansion of any of the $f_i(\vec{\sigma})$.

Again following \cite{Steinacker_2010}, the natural Laplace operator is given by
\begin{equation}\label{eqn:nc-lap}
\Delta f(\vec{x}) := \sum_{i=1}^D[X^i,[X^i,f(X^i)]].
\end{equation}
In the commutative limit (meaning to leading order in $\theta$ so we don't have to worry about ordering in multiplication), this operator may be written in terms of some metric $G^{ab}$ as $\Delta = \frac{1}{\sqrt{G}}\partial_a(\sqrt{G}G^{ab})\partial_b = G^{ab}\nabla_a \nabla_b$. $G^{ab}$ is determined in terms of $\theta$ and the constraint functions $h^r(X)$ by the relations 
\begin{equation}\label{eqn:G-def}
\begin{split}
&\sqrt{G}G^{ab}(\vec{\sigma}) := \frac{1}{\sqrt{\theta}}\theta^{aa'}(\vec{\sigma})\theta^{bb'}(\vec{\sigma})g_{a'b'}(\vec{\sigma}),\\
&g_{ab}(\vec{\sigma}) := \partial_a X^i \partial_b X^j\delta_{ij}.
\end{split}
\end{equation}
$g_{ab}$ is the induced metric on $\mathcal{M}_{\theta}$ when the metric in $\mathbb{R}^D_{\theta}$ is $\delta_{ij}$. To see that the volume form $\sqrt{G(\vec{\sigma})}d^d\sigma$ must indeed agree with $\mu(\vec{\sigma})d^d\sigma$ in \eqref{eqn:nc-int}, we note that $\Tr[\Delta f(\vec{x})]$ for the Laplacian in \eqref{eqn:nc-lap} always vanishes, as the trace of a commutator on a finite dimensional vector space must always vanish. Therefore, the quantity
\begin{equation}
\mu(\vec{\sigma})\Delta f(\vec{\sigma}) = \frac{\mu(\vec{\sigma})}{\sqrt{G(\vec{\sigma})}}\partial_a((\sqrt{G}G^{ab})\partial_b f(\vec{\sigma}),
\end{equation}
must be a total derivative for all $f(\vec{\sigma}) \in \mathfrak{a}_{\mathcal{M}_{\theta}}$, so $\mu/\sqrt{G}$ must be constant over $\mathcal{M}_{\theta}$.

When we consider MQM it is precisely the representation matrices $X^i$ that will become our dynamical degrees of freedom. As such, the natural metric $G^{ab}$ on $\mathcal{M}_{\theta}$ will also be dynamical.

\subsection{Target Space Entanglement and MQM Edge Modes}\label{ssec:rev-ent}

We review the construction of \cite{Hartnoll:2015fca,Han:2019wue,Das:2020jhy,Hampapura:2020hfg}, and its synthesis with the notion of Gauss law entanglement as defined in \cite{Frenkel:2021yql,Frenkel:2023aft}. For more general introductions to entanglement in gauge theories, see \cite{Ghosh:2015iwa,Soni:2015yga, Donnelly:2016auv}. 

\begin{figure}[h]
\begin{center}
\includegraphics[width=0.6 \textwidth]{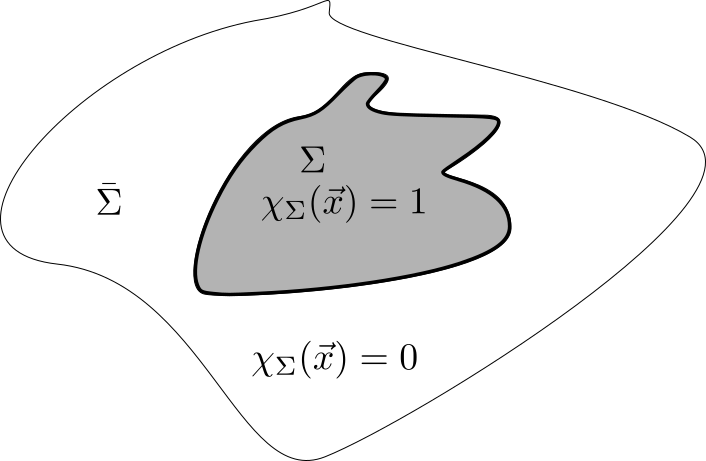}
\caption{A depiction of the characteristic function $\chi_{\Sigma}(\vec{x})$ associated to a subset of a manifold.} \label{fig:char-func}
\end{center}
\end{figure}

Consider a commutative manifold $\mathcal{M}$ parameterized by coordinates $\vec{x}$. A subregion $\Sigma$ of $\mathcal{M}$ is defined via a characteristic function $\chi_{\Sigma}(\vec{x})$, which takes values 1 inside $\Sigma$ and 0 in its complement $\bar{\Sigma}$ (see Fig. \ref{fig:char-func}).

If we have some fields $\phi_{\alpha}$ (with their associated momenta $\pi_{\alpha}$) propagating on $\mathcal{M}$, the degrees of freedom associated to a subregion are generated by $\chi_{\Sigma}(\vec{x})\phi_{\alpha}(\vec{x})$ and $ \chi_{\Sigma}(\vec{x})\pi_{\alpha}(\vec{x})$\footnote{There are a lot of subtleties related to defining these subalgebras for continuum field theories (see \cite{Witten:2018zxz, Casini:2022rlv} for discussion on this point). We avoid these subtleties as we are working with a finite number of degrees of freedom (a type I von Neumann algebra), so we do not go into further detail here.}. 

To instead partition the target space of an MQM theory associated to $\mathcal{M}_{\theta}$, we promote this characteristic function to have the matrices $X^i$ as arguments and define the matrix $\Theta_{\Sigma}$ as
\begin{equation}\label{eq:thet-sig}
    \Theta_{\Sigma} := \chi_{\Sigma}(\vec{X}).
\end{equation}
$\Theta_{\Sigma}$ is generically a projection matrix in the sense that $\Theta_{\Sigma}^2 = \Theta_{\Sigma}$. To see this, we define $\Sigma$ as the interior of the level set $f(\vec{x}) = c$ of some real coordinate function $f(\vec{x})$: 
\begin{equation}
    \chi_{\Sigma}(\vec{x}) = \theta(f(\vec{x}) - c).
\end{equation}
Here, $\theta$ is the Heaviside step function. $f$ is chosen to be real, so it may be promoted to a hermitian matrix\footnote{Due to ordering ambiguities, we must make an explicit ordering choice to ensure a real coordinate function $f(\vec{x})$ maps to a unitarily diagonalizable matrix. Corrections to the leading geometric term in the entanglement from ambiguities in this choice are subleading in $\theta$ for entanglement cuts whose radius of curvature is much larger than the noncommutativity lengthscale.} $F$ in the representation of the noncommutative algebra, with eigenvalues $f_n$. We evaluate $\Theta_{\Sigma}$ in the basis where $F$ is diagonal -
\begin{equation}
    \Theta_{\Sigma} = \begin{bmatrix}
    \theta(f_1 - c) & 0 & \ldots & 0\\
    0 & \theta(f_2 - c) & \ldots & 0\\
    \vdots & \vdots & \ddots & \vdots\\
    0 & 0 & \ldots & \theta(f_N - c)
    \end{bmatrix}.
\end{equation}
$\theta(f_i - c)$ will take on values 1 or 0 as long as $f_i \neq c$. Matrices that have an eigenvalue precisely equal to $c$ make up a measure zero subset of the configuration space, so however we choose to define the theta function value at $f_i = c$ will have no effect on expectation values of observables\footnote{It is this sense in which noncommutative field theories emergent from MQM regularize the ill-behaved quadratic forms associated to the boundary values of the fields $\phi_{\alpha}(f(\vec{x}) = c)$ (see the discussion in section 4 of \cite{Penington:2023dql}).}.

The matrices naturally decompose into four blocks
\begin{equation}\label{eq:Mblocks}
    X^i_{AB}:=\Theta_{A}X^i\Theta_{B},\quad A,B \in \{\Sigma,\bar{\Sigma}\},
\end{equation}
and in the basis where $\Theta_{\Sigma}$ is diagonal we may associate each of these matrices to sub-blocks in a block decomposition (see Fig. \ref{fig:block-decomp}). We note a peculiarity of this partition - the conjugate momentum to $X^i_{\Sigma \Sigma}$ (denoted $\Pi^i_{\Sigma \Sigma}$) is \textit{not} equal to $\Theta_{\Sigma}\Pi^i\Theta_{\Sigma}$. This statement would only hold if we were to gauge fix to break the overall $SU(N)$ symmetry group down to the $U(M)\cross U(N-M)$ subgroup that commutes with $\Theta_{\Sigma}$. One sees this by considering the kinetic term of the Lagrangian
 \begin{equation}
 \begin{split}
     &L_K = \Tr[\dot{X}^{i2}] = \Tr[[(\dot{X}^i_{\Sigma \Sigma} + \dot{X}^i_{\Sigma \bar{\Sigma}} + \dot{X}^i_{\bar{\Sigma}\Sigma} + \dot{X}^i_{\bar{\Sigma}\bar{\Sigma}})]^2]=\\
     &=\Tr[\dot{X}^i_{\Sigma \Sigma}\left(\dot{X}^i_{\Sigma \Sigma} - \dot{\Theta}_{\Sigma}X^i\Theta_{\Sigma} - \Theta_{\Sigma}X^i\dot{\Theta}_{\Sigma}\right) + \ldots],
\end{split}
\end{equation}
where the second line follows from direct computation by noting $\Theta_{\Sigma} = \text{Id}_N - \Theta_{\bar{\Sigma}}$. We therefore have, for example,
\begin{equation}
    \Pi^i_{\Sigma \Sigma} = \dot{X}^i_{\Sigma \Sigma} - \dot{\Theta}_{\Sigma}X^i\Theta_{\Sigma} - \Theta_{\Sigma}X^i\dot{\Theta}_{\Sigma}.
\end{equation}
This dressing of the conjugate momentum stems from the fact that generically $\Theta_{\Sigma}\dot{\Theta}_{\bar{\Sigma}} \neq 0$, allowing for kinetic coupling between $\dot{X}^i_{\Sigma \Sigma}$ and the $\Sigma \bar{\Sigma},\bar{\Sigma}\Sigma$ blocks.

Wavefunctions on the $\Sigma \Sigma$ degrees of freedom may be expressed as $\Psi(X^i_{\Sigma \Sigma}) = \Psi(X^i_{\Sigma \Sigma,gf},U)$. The $X^i_{\Sigma \Sigma,gf}$ are a choice of variables along some gauge-fixing slice, and $U$ parameterizes the $U(M)$. We may therefore write the $\Sigma \Sigma$ Hilbert space as
\begin{equation}
\mathcal{H}_{\Sigma \Sigma} = \mathcal{H}_{gf} \otimes \mathcal{H}_{U(M)}.
\end{equation}
Via the Peter-Weyl theorem, we may write $\mathcal{H}_{U(M)}$ as a direct sum over all irreducible representations of $U(M)$, which we label by $r$:
\begin{equation}
\mathcal{H}_{U(M)} = \bigoplus_r \mathcal{H}_{r}.
\end{equation}
The generator of the $SU(N)$ global symmetry is
\begin{equation}
G := 2i[X^i,\Pi^i].
\end{equation}
Whether or not we take this symmetry to be gauged, we will be interested in singlet states under $U(N)$. These are states $\ket{\Psi}$ that satisfy
\begin{equation}
G_{nn'}\ket{\Psi} = 0 \quad \forall n,n'.
\end{equation}
We can apply the block diagonal decomposition, in particular, to the Gauss law to write
 \begin{equation}
     G_{\Sigma \Sigma} = 2i\sum_i\left([X^i_{\Sigma \Sigma},\Pi^i_{\Sigma \Sigma}] + X^i_{\Sigma \bar{\Sigma}}\Pi^i_{\bar{\Sigma}\Sigma} - \Pi^i_{\Sigma \bar{\Sigma}}X^i_{\bar{\Sigma \Sigma}}\right).
 \end{equation}
The constraint $G_{\Sigma \Sigma}\ket{\Psi} = 0$ entangles the $\Sigma \Sigma$ and $\Sigma \bar{\Sigma},\bar{\Sigma}\Sigma$ degrees of freedom, and it is this contribution to the entanglement that we refer to as the Gauss law entanglement. In particular, it is useful to define the objects
\begin{equation}\label{eqn:Q-sig-def}
    \begin{split}
        G_{\Sigma}:=2i\sum_i[X^i_{\Sigma \Sigma},\Pi^i_{\Sigma \Sigma}], \quad &Q_{\Sigma}:=2i\sum_i\left(\Pi^i_{\Sigma \bar{\Sigma}}X^i_{\bar{\Sigma}\Sigma} - X^i_{\Sigma \bar{\Sigma}}\Pi^i_{\bar{\Sigma}\Sigma}\right),\\
        G_{\Sigma}\ket{\Psi} &= Q_{\Sigma}\ket{\Psi}.
    \end{split}
\end{equation}

$G_{\Sigma}$ is the natural set of $U(M)$ generators one can construct out of the $\Sigma \Sigma$ degrees of freedom, and likewise $Q_{\Sigma}$ for the off-diagonal block. The second line in \eqref{eqn:Q-sig-def} implies the reduced density matrix on the $\Sigma \Sigma$ degrees of freedom can (and generically will) have support on nontrivial irreps of the $SU(M)$ group generated by $G_{\Sigma}$. These irreps are fixed to be identical to the representations under $Q_{\Sigma}$ carried by the off-diagonal blocks. When promoted to quantum operators, the matrix elements of $G$ commute with $\ket{\Psi}\bra{\Psi}$. Upon tracing out the $\Sigma \bar{\Sigma},$ $\bar{\Sigma}\Sigma$, and $\bar{\Sigma}\bar{\Sigma}$ degrees of freedom we therefore have
\begin{equation}\label{eqn:G-commute}
    [[G_{\Sigma \Sigma,nn'},\rho_{\Sigma}]]_q = 0 \implies [[G_{\Sigma,nn'},\rho_{\Sigma}]]_q = 0,
\end{equation}
where $[[\cdot,\cdot]]_q$ denotes the commutator of the matrix elements as quantum operators. This notation is to distinguish the commutator of operators on the Hilbert space from commutators of the classical matrices. The implication is deduced from the fact that the $Q_{\Sigma}$ is built out of operators that do not act on the $\Sigma \Sigma$ degrees of freedom.

\eqref{eqn:G-commute} together with Schur's lemma now tells us that $\rho_{\Sigma}$ decomposes into a direct sum of operators proportional to the identity on each irrep subspace of $\mathcal{H}_{U(M)}$:
\begin{equation}
    \rho_{\Sigma} = \bigoplus_r \rho_r, \quad \rho_r = \frac{p_r}{\dim r}\mathbb{1}_r \otimes \rho_{r,gf}.
\end{equation}
$p_r$ is the probability of being in the irrep $r$, $\dim r$ is the dimension of $\mathcal{H}_r$, $\mathbb{1}_r$ is the identity matrix on $\mathcal{H}_r$, and $\rho_{r,gf}$ is the state of the remaining degrees of freedom on $\mathcal{H}_{gf}$, which generically depends on $r$. We now simply calculate the von Neumann entropy to find
\begin{equation}\label{eqn:vN-entr}
-\Tr[\rho_{\Sigma} \log \rho_{\Sigma}] = \sum_r p_r \log \dim r - p_r \log p_r - \Tr[\rho_{r,gf}\log \rho_{r,gf}].
 \end{equation}

The first term is the expectation value of the irrep dimension. It is this term that we focus on in this note. The second term is a classical Shannon entropy over the superselection sectors, and the third term is the entanglement due to the singlet fluctuations within each superselection sector. We interpret the work of \cite{Karczmarek:2013jca,Han:2019wue} as having calculated the contribution due to $\rho_{r,gf}$. In local gauge field theories, it is $\langle \log \dim r \rangle$ that dominates the entanglement and contributes to the leading area law divergence. In electromagnetism coupled to matter, for example, $r$ is set by the number of virtual electron-positron pairs straddling the entanglement cut.

For the $U(N)$ groups we will be considering, irreps are labeled by Young diagrams. Our goal will be to determine the Young diagram that dominates the sum in \eqref{eqn:vN-entr}.

\subsubsection{Edge Mode Irreps in the Classical Limit}\label{sssec:class-limit}

\begin{figure}[h]
    \centering
    \includegraphics[width=0.5\textwidth]{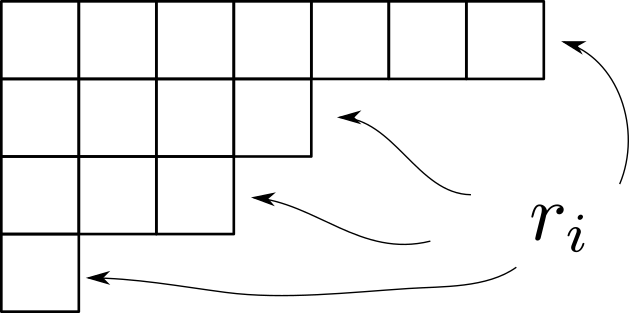}
    \caption{The entanglement edge mode states we find are maximally mixed over irreducible representations of $U(M)$. Such irreps are labeled by Young diagrams. }
    \label{fig:gen-yt}
\end{figure}

For the rest of this note we assume that the MQM wavefunction localizes to a classical minimum
\begin{equation}
X^i = X^i_{cl} + \delta X^i = X^i_{cl} + \sum_a \delta x_a Y^i_a.
\end{equation}
The $\delta x_a$ are perturbations with conjugate momentum $\delta \pi_a$ and the $Y^i_a$ are the quadratic normal modes of the (possibly fully loop corrected) action around $X_{cl}^i$ and satisfy
\begin{equation}
\sum_i \Tr[Y_a^{i \dag}Y_b^i] = \delta_{ab}.
\end{equation}
We denote the characteristic frequency of $Y_a^i$ as $\omega_a$. To leading order in $\delta x_a$ we may write
\begin{equation}
\begin{split}
&X^i = X^i_{cl},\quad \Pi^i = \sum_a \delta \pi_a Y^i_a \implies\\
&Q_{\Sigma} = 2i \sum_i \sum_a \delta \pi_a \left(Y^i_{a,\Sigma \bar{\Sigma}}X^i_{cl, \bar{\Sigma}\Sigma} - X^i_{cl,\Sigma \bar{\Sigma}}Y^i_{a,\bar{\Sigma}\Sigma}\right).
\end{split}
\end{equation}
In this limit we note the strange feature that the elements of $Q_{\Sigma}$ all commute amongst themselves. This removes the typical ordering ambiguity in defining the higher Casimir operators $\Tr[Q_{\Sigma}^{2p}]$. Because $\Tr[Q_{\Sigma}^{2p}]$ is constant over irreps, we may easily evaluate its action of on any irrep by considering the highest weight state $\ket{h.w.}$. This is the state with the Young tableau labeled with all 1s on the first row, all 2s on the second, etc. Because the highest weight state is annihilated by all $Q_{\Sigma,nn'}$ with $n < n'$, we immediately have
\begin{equation}\label{eqn:cas-hw}
\Tr[Q_{\Sigma}^{2p}]\ket{h.w.} = \sum_i r_i^{2p}\ket{h.w.},
\end{equation}
where $r_i$ is the number of boxes in the $i$th row (see Fig. \ref{fig:gen-yt}).

Lastly we make the comment that for the cases we consider, the off-diagonal blocks of the classical minima $X^i_{cl,\bar{\Sigma}\Sigma}$ will be low-rank. The intuition for this is that off-diagonal elements of the matrices $x^i_{nn'}$ represent the states of strings stretched between the $n$th and $n'$th $D0$-branes. Most pairs of $D0$-branes will be separated on the order of the lengthscale of the geometry, so most $x^i_{nn'}$ will be suppressed to be near zero in the ground state except those stretching between branes near the entangling surface.

This means that for $p$ that doesn't scale with $N$ the leading in $N$ contribution to $\Tr[Q_{\Sigma}^{2p}]$ will come from terms where we collect the $Y^i$ together, so we may write
\begin{equation}\label{eqn:Q-sig-gen}
\Tr[Q_{\Sigma}^{2p}] = \sum_{\{i\}} \Tr[X_{cl}^{i_1}\ldots X_{cl}^{i_k}\Pi^{i_{1}}\ldots \Pi^{i_k}] + O(1/N).
\end{equation}

In particular, this means the matrix $Q_{\Sigma}$ itself will be low rank on account of it having a factor of $X_{cl}^i$. This fact and \eqref{eqn:cas-hw} imply to leading order in $N$ and $\theta$, the eigenvalues of $Q_{\Sigma}^2$ are given by $r_i^2$ - the squares of each row length. This will allow us to read off the dominant irrep in the sum \eqref{eqn:vN-entr} by considering the expectation values $\left\langle \Tr[Q_{\Sigma}^{2p}]\right\rangle$ in the state of interest.

Given the shape of the Young diagram in Fig. \ref{fig:gen-yt} (and remembering that it it is an irrep of $U(M)$), the entropy is now given by evaluating $\log \dim r$, which is directly evaluated in two different regimes:
\begin{equation}\label{eqn:ent-from-yt}
\begin{split}
&\log \dim r = \sum_i r_i \log(M/r_i) + O(r_i), \quad r_i \ll M,\\
&\log \dim r = \sum_i M\log(r_i/M) + O(M), \quad r_i \gg M. 
\end{split}
\end{equation}
We will find that the row lengths $r_i$ are given by the surface areas of different topologically disconnected parts of $\partial \Sigma$. Remembering that $M$ is the volume of $\Sigma$, the two regimes in \eqref{eqn:ent-from-yt} represent a transition from a boundary area-law regime to a volume-law regime. 
 
\section{The Fuzzy Sphere}\label{sec:fuzzsphere}

\begin{figure}[h]
    \centering
    \includegraphics[width=0.65\textwidth]{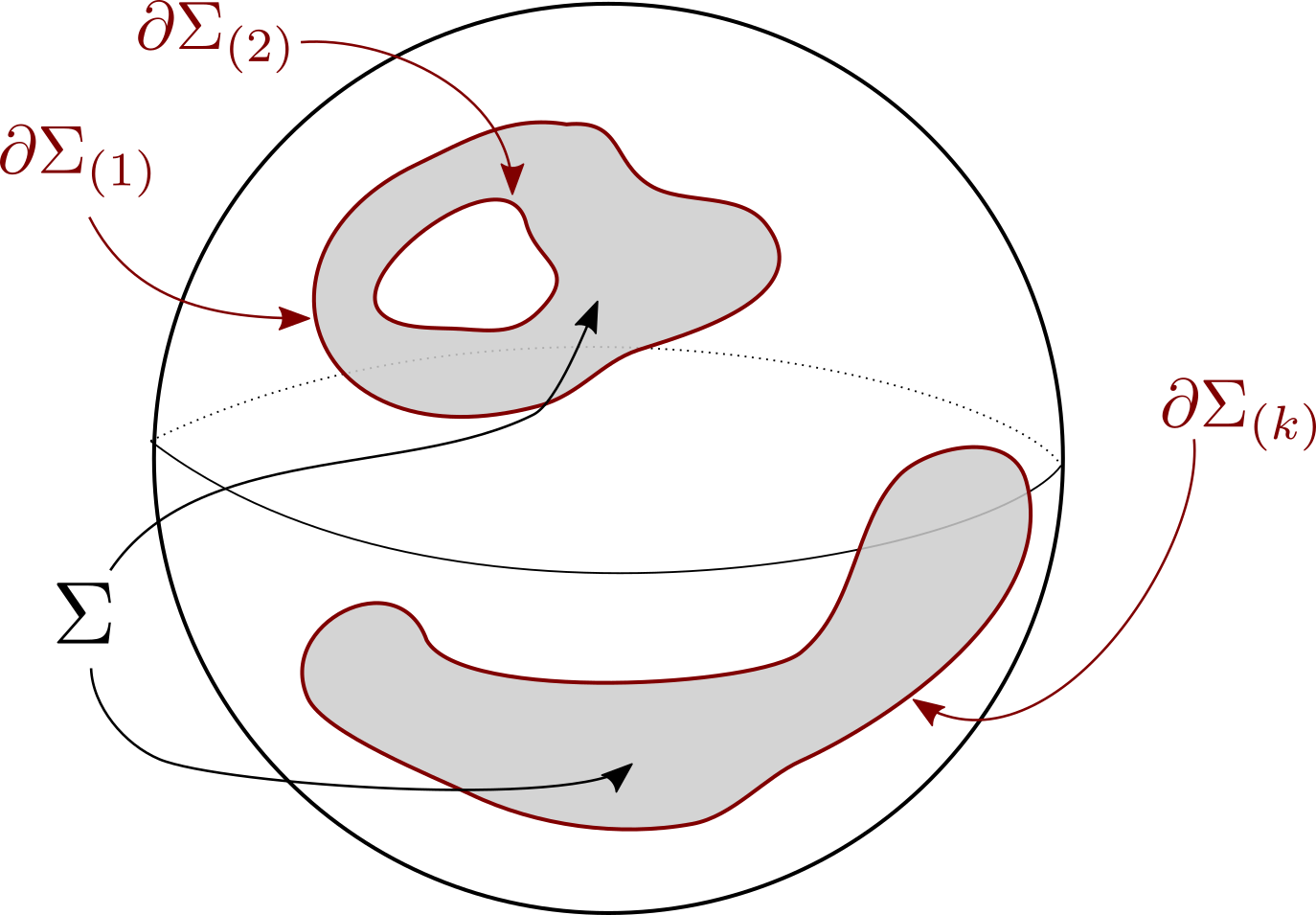}
    \caption{A possible entanglement cut drawn on the fuzzy sphere. The subregion of interest, $\Sigma$, is shaded in gray. It may have any topology, and have multiple disconnected regions that themselves do not need to be simply connected. We label each disconnected component of the entanglement cut by $\partial \Sigma_{(k)}$. In this diagram, $\Sigma$ has two components but $\partial \Sigma$ has three components.}
    \label{fig:multicut}
\end{figure}

We first consider the fuzzy sphere \cite{Madore:1991bw}, extending the results of \cite{Frenkel:2023aft} to arbitrary subregions and nonabelian emergent gauge fields. We only consider the edge mode contribution to the entanglement - for the bulk contribution, see \cite{Han:2019wue,Karczmarek:2013jca}. We consider entanglement cuts of all possible configurations and topologies, such as that depicted in Fig. \ref{fig:multicut}, so long as their radius of curvature is parametrically larger than the noncommutativity scale.
The model we consider is one with three bosonic $N \times N$ matrices $X^i$ with lagrangian
\begin{equation}\label{eqn:mini-bmn}
    L = \Tr[\dot{X}^{i2} + (i\nu X^k + \epsilon^{ijk}[X^i,X^j])^2].
\end{equation}
The square structure of the potential imposes the $su(2)$ algebra upon the classical minima $X_{cl}$ in the configuration space of matrices:
\begin{equation}
    i\epsilon^{ijk}[X_{cl}^i,X_{cl}^j] = \nu X_{cl}^k \implies X_{cl}^i = \nu J^i.
\end{equation}
Here, $J^i$ are generators of an $N$-dimensional representation of $SU(2)$. There is a host of classical minima, each associated to an $N$-dimensional representation of $\mathfrak{su}(2)$. This representation may be reducible or irreducible, although only the ground states corresponding to irreducible representations correspond to exact quantum ground states in the supersymmetric model \cite{Han:2019wue}. For the purely bosonic theory, only the trivial minimum at $X_{cl}^i = 0$ corresponds to the unique quantum ground state.

As in \cite{Han:2019wue,Frenkel:2023aft}, we consider the semiclassical approximation to the wavefunction localized to the classical ground state. This means we expand the dynamical fields $X^i$ around $X^i_{cl}$ as
\begin{equation}
    X^i = X^i_{cl} + \sum_{a} \delta x_a Y^i_a, \quad Y^i_a = \sum_{jm}y^i_{a,jm}\hat{Y}_{jm}.
\end{equation}
The $Y^i_a$ are the normal modes around the classical minimum solved for in Appendix C of \cite{Han:2019wue}. The normal modes satisfy
\begin{equation}\label{eqn:fuzzy-eom}
Y_a^i + i\epsilon^i_{jk}[J^i,Y_a^k] = \omega_aY_a^i, \quad \sum_i \Tr[(Y_a^i)^{\dag}Y_b^i] = \delta_{ab}.
\end{equation}

\subsection{$U(1)$ Backgrounds}\label{ssec:u1}

We focus first on the irreducible representations, the low energy fluctuations around which correspond to an emergent noncommutative electrodynamics on the sphere (see appendix A of \cite{Han:2019wue}). Around this minimum, an emergent spherical geometry is defined precisely as in sec. \ref{ssec:rev-ncb}. We generate the algebra of functions by associating
\begin{equation}\label{eq:nc-angs}
    X_{cl}^i \leftrightarrow x^i, \quad i[X_{cl}^i,\cdot] \leftrightarrow \tilde{L}_i,
\end{equation}
where the $\tilde{L}_i$ are noncommutative angular momentum operators, and approach the usual commutative momentum generators $L_k = i\epsilon^{ijk}x^i\partial_j$ in the large $N$ limit. Exponentiating the commutator in \eqref{eq:nc-angs}, we find a natural embedding of the global $SO(3)$ Killing symmetry of the sphere into the $U(N)$. Parameterizing group elements $g \in SO(3)$ by the Euler angles $g = e^{i \alpha \Lambda_3}e^{i \beta \Lambda_1}e^{i \gamma \Lambda_3}$ we have
\begin{equation}
SO(3) \hookrightarrow U(N), \quad g \rightarrow U(g):=e^{i \alpha X_{cl}^3}e^{i \beta X_{cl}^1}e^{i \gamma X_{cl}^3}.
\end{equation}

\begin{figure}[h]
\begin{center}
\includegraphics[width=0.7\textwidth]{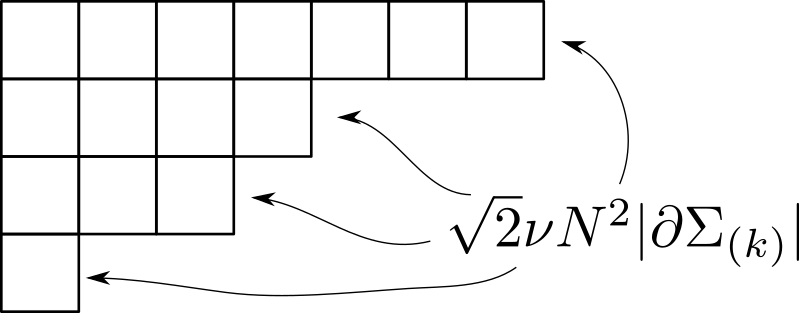}
\caption{The Young diagram of the irrep dominating the state of the edge modes for an entanglement cut on the fuzzy sphere. There are as many rows in the diagram as topologically disconnected pieces of the entanglement cut. Each row length $r_k$ is given by the length of one of the disconnected pieces cut $|\partial \Sigma_{(k)}|$ pictured in Fig. \ref{fig:multicut}. This area-law behavior emerges despite the UV/IR mixing endemic to noncommutative geometries (see e.g. \cite{Karczmarek:2013jca}). We have normalized the fuzzy sphere to have unit radius, so the row lengths scale with $N^2$.}
\label{fig:fuzz-sphere-yt}
\end{center}
\end{figure}

Our task is now to apply the technology of \S\ref{ssec:rev-ent} to an entanglement cut such as that pictured in Fig. \ref{fig:multicut}. In particular, we would like to understand the shape of the Young diagram that dominates expectation values of $\Tr[Q_{\Sigma}^{2p}]$. Our results are quite simply stated with respect to Fig. \ref{fig:gen-yt} - each row length $r_i$ of the diagram is proportional to the length of one of the disconnected components $|\partial \Sigma_{(k)}|$ of the entanglement cut, scaling with powers of $\nu$ and $N$ (see Fig. \ref{fig:fuzz-sphere-yt}). 

The technical details of this calculation are reserved for Appendix \ref{app:fuzz-sph}. The important result is that up to $O(1/N)$ corrections, $\langle \Tr[Q_{\Sigma}^2] \rangle$ is given by a matrix element of the operator $\mathcal{O}_{Q}(\cdot):= \sum_{ij}K_{ij}[J^i,[J^j,\cdot]]$ on the space of functions on the fuzzy sphere. 
\begin{equation}\label{eqn:OQ-def}
\Tr[Q_{\Sigma}^{2}] = \Tr[\Theta_{\Sigma}\mathcal{O}_{Q}(\Theta_{\Sigma})] = \sum_{ij}\Tr[\Theta_{\Sigma}K_{ij}[J^i[J^i,\Theta_{\Sigma}]]]. 
\end{equation}
In particular, the functions $K_{ij}$ represent slowly-varying\footnote{When we use the term slowly-varying in this note, we always mean with respect to the noncommutativity scale (which also functions as the UV cutoff).} functions on the fuzzy sphere, in the sense that the magnitudes of the entries of their derivatives $[J^k,K_{ij}]$ are comparable with those of $K_{ij}$ itself for all $i,j,k$. They are calculated as
\begin{equation}
K_{ij} = \langle \Pi^i \Pi^j \rangle \approx \sum_{a}\omega_a Y^i_a Y^j_a + O(1/\nu N), \quad K_{ij} \leftrightarrow \kappa_{ij}(\theta,\phi).
\end{equation}

This sum is dominated by high energy modes whose frequency is of the order of the UV cutoff, so the fact that they are smooth slowly varying functions is a priori surprising. Because they are slowly varying, $\mathcal{O}_Q$ has the interpretation of a smooth second order differential operator on the fuzzy sphere under the Moyal map:
\begin{equation}
\sum_{ij}K_{ij}[J^i,[J^j,\cdot]] \leftrightarrow \sum_{ij}\kappa_{ij}(\theta,\phi)L_iL_j.
\end{equation}
Because it is calculated from an $SO(3)$ invariant ground state, this operator must respect the symmetries of the fuzzy sphere, in the sense that
\begin{equation}
\Tr[UF'U^{\dag}\mathcal{O}_Q(UFU^{\dag})] \quad 
 \forall F,F'\in \mathfrak{a}_{S_{\theta}^2}, \forall U \in SO(3) \subset U(N).
\end{equation}

This implies, again by Schur's lemma, that $\mathcal{O}_Q$ must decompose as a sum over operators proportional to the identity on the irreducible representations such that
\begin{equation}
\mathcal{O}_Q(\hat{Y}_{jm}) = f(j)\hat{Y}_{jm} \quad \forall \hat{Y}_{jm},
\end{equation}
for some function $f(j)$ of the $SO(3)$ irrep label $j$. To make progress, we expand $\Theta_{\Sigma}$ in terms of spherical harmonics
\begin{equation}\label{eqn:thet-decomp}
    \Theta_{\Sigma} = \sum_{jm}c_{\Sigma,jm}\hat{Y}_{jm}, \quad c_{\Sigma,jm} = \Tr[\Theta_{\Sigma}\hat{Y}_{jm}^{\dag}]=\int_{S^2} \sqrt{g}d\theta d\phi\, \chi_{\Sigma}(\theta,\phi)y_{jm}(\theta,\phi).
\end{equation}
The $c_{\Sigma,jm}$s are therefore precisely the coefficients of $\chi_{\Sigma}$ in the usual commutative spherical harmonic expansion. They may be calculated directly by noting for large $j$ (when harmonics are locally well approximated by plane waves) they are simply the Fourier decomposition of the step function, and so satisfy
\begin{equation}
    \sum_m |c_{\Sigma, jm}|^2 = \sum_k \frac{|\partial \Sigma_{(k)}|^2}{j^2} + O(1/j^3) \implies \langle \Tr[Q_{\Sigma}^2] \rangle = \sum_k |\partial \Sigma_{(k)}|^2 \sum_j \frac{f(j)}{j^2}.
\end{equation}

Putting everything together, and setting the overall coefficient $\sum_j\frac{f(j)}{j^2} = 2\nu^2N^4$ by the results of \cite{Frenkel:2023aft} for the cap subregion, we therefore have
\begin{equation}
\Tr[Q_{\Sigma}^2] \approx 2\nu^2N^2\sum_i\Tr[\Theta_{\Sigma}[J^i,[J^i,\Theta_{\Sigma}]]] \approx \sum_{k}2\nu^2N^4|\partial \Sigma_{(k)}|^2.
\end{equation}
More generally, due to their delta function structure the cross terms between topologically disconnected pieces of the entanglement cut vanishes to leading order in $N$. This property allows us to calculate
\begin{equation}
\Tr[Q_{\Sigma}^{2p}] = \sum_k 2^p\nu^{2p} N^{4p}|\partial \Sigma_{(k)}|^{2p}.
\end{equation}
The corrections will depend on the curvature of the entanglement cut and the separation between disconnected pieces. What we may deduce from this is that the eigenvalues of $Q_{\Sigma}$ are precisely $\sqrt{2} \nu N^2 |\partial \Sigma_{(k)}|$ - each eigenvalue precisely given by the length of a topologically disconnected piece of the entanglement cut (Fig. \ref{fig:fuzz-sphere-yt}).

\subsection{$U(q)$ Backgrounds}\label{ssec:uq}

To arrive at a $U(q)$ emergent non-abelian gauge theory on the fuzzy sphere, we instead expand around the classical background given by
\begin{equation}\label{eq:nacl}
X^i_{cl} = \nu J^i \otimes \mathbb{1}_q.
\end{equation}

For irreducible generators $J^i$. We take $J^i$ to be an $N \times N$, so each $X^i_{cl}$ is an $Nq \times Nq$ matrix. One can think of this state as having $q$ fuzzy spheres stacked on top of one another, and the $U(q)$ symmetry transforms these spheres into each other. These backgrounds will not correspond to a ground state of \eqref{eqn:mini-bmn}, but will still correspond to metastable local minima of the one-loop bosonic action. The equations of motion will be identical to \eqref{eqn:fuzzy-eom} with the irreducible generators replaced by the reducible generators in \eqref{eq:nacl}:
\begin{equation}\label{eqn:na-eom}
Y_{q,a}^i + \epsilon^i_{jk}[J^i \otimes \mathbb{1}_q,Y_{q,a}^k] = \omega_a Y_{q,a}^i.
\end{equation}

We label the normal modes for $q > 1$ as $Y_{q,a}^i$ to distinguish them from the $U(1)$ normal modes in the previous section, which we continue to label $Y_{a}^i$. At this order in perturbation theory, therefore, any matrix of the form $Y_{q,a}^i = Y_a^i \otimes W$ for any $q \times q$ matrix $W$ will solve the equations of motion, as any $W$ will commute with the identity in \eqref{eqn:na-eom}. This gives us a basis of $N^2q^2$ solutions to the equations of motion, which means we have found a complete basis. Because we may choose any basis of $W$, it directly follows that the $K_{ij}$ matrices in $\mathcal{O}_Q$ will now be given as
\begin{equation}
K_{q,ij} = K_{ij}\otimes \mathbb{1}_q.
\end{equation}

Where again we use a subscript $q$ on $K_{ij}$ to distinguish the $q>1$ case from the $q=1$ case. We repeat the same calculations, but each topologically disconnected piece of the entanglement cut contributes $q$ rows instead of 1. The reason for this may again be seen by considering the cap subregion, where $J^i_{\bar{\Sigma}\Sigma}$ is now rank $q$ instead of rank $1$. To leading order (if $q$ doesn't scale with $N$) we again have
\begin{equation}
\Tr[Q_{\Sigma}^{2p}] = \sum_k q2^p\nu^{2p} N^{4p}|\partial\Sigma_{(k)}|^{2p}.
\end{equation}

\begin{figure}[h]
\begin{center}
\includegraphics[width=0.65\textwidth]{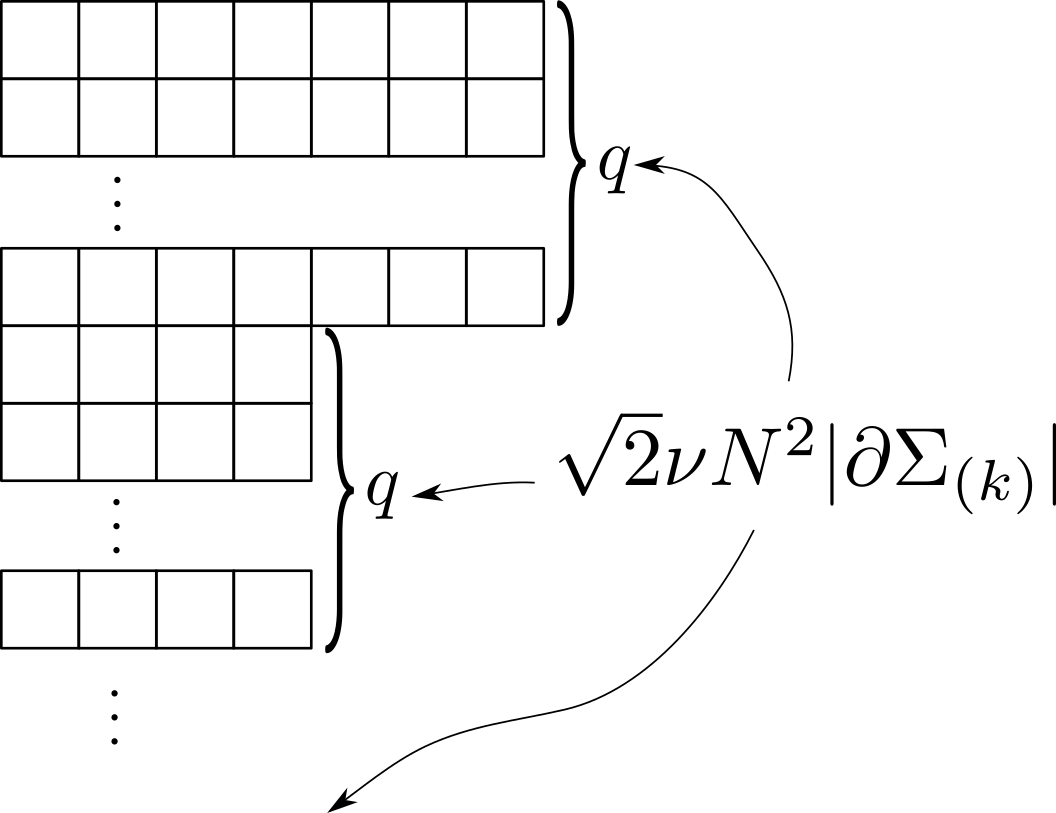}
\caption{The shape of the dominant Young diagram for emergent $U(q)$ backgrounds on the fuzzy sphere. The only difference from the $U(1)$ case is that each row length is repeated $q$ times, so there are $q$ times as many rows in the diagram as there are topologically disconnected pieces of the entanglement cut.}
\label{fig:fuzz-sphere-yt-2}
\end{center}
\end{figure}

The shape of the dominant Young diagram is easy to deduce -- instead of one row for every disconnected piece of the entanglement cut, there are be $q$ rows of length $\sqrt{2}\nu N^{2}|\partial \Sigma_{(k)}|$ (see Fig. \ref{fig:fuzz-sphere-yt-2}). It is interesting to point out that in these Young diagrams the $U(q)$ symmetry of the edge modes is manifest. For Yang-Mills on commutative manifolds, entanglement edge modes transform in the fundamental representation of $U(q)$ \cite{Donnelly:2016auv}. This behavior is also manifest here - the emergent $U(q)$ symmetry takes the different rows of equal length in Fig. \ref{fig:fuzz-sphere-yt-2} into each other.

To leading order in $N,\nu$, the contribution to the entanglement is simply enhanced by a factor of $q$ with respect to the $U(1)$ case, as can be seen directly from \eqref{eqn:ent-from-yt}. This is entirely expected when comparing to the commutative limit, as fundamental $U(q)$ edge modes roughly have $q$ times as many degrees of freedom, so their entropy should be correspondingly enhanced by a factor of $q$. 

\section{Free Fields on General Noncommutative Backgrounds}\label{sec:gen-mani}

\begin{figure}[h]
    \centering
    \includegraphics[width=0.65\textwidth]{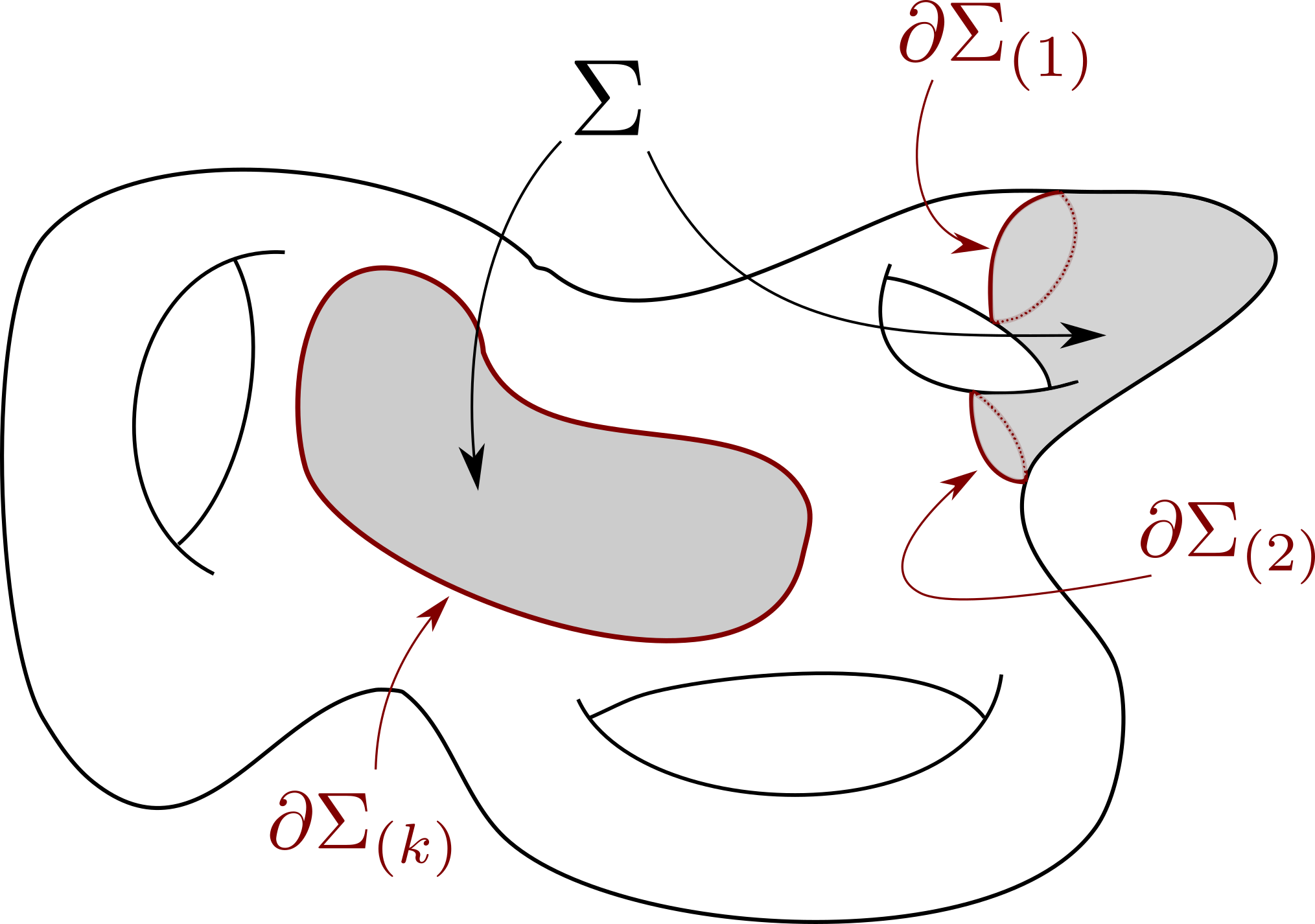}
    \caption{We consider entanglement on some general (compact without boundary) noncommutative manifold $\mathcal{M}_{\theta}$, defined via some embedding into $\mathbb{R}^D_{\theta}$ as in \cite{steinacker2011non}. On such manifolds, even the dynamics of free particles are typically chaotic. See e.g. \cite{applebaum1988stochastic} for an explicit example of dynamics on a noncommutative two-torus.}
    \label{fig:gen-mani}
\end{figure}

We comment on how the above results generalize to arbitrary noncommutative backgrounds $\mathcal{M}_{\theta}$ (e.g. that depicted in Fig. \ref{fig:gen-mani}). These objects may in general have no global symmetries and complicated topology -- the only assumption we make is that their scale of curvature is much larger than their scale of noncommutativity.

In general, even a free particle propagating on a commutative Riemmanian manifold with no global symmetries will behave chaotically. Its energy eigenstates will obey the eigenstate thermalization hypothesis (ETH) \cite{srednicki1994chaos,deutsch2018eigenstate} and Berry's conjecture \cite{berry1977regular} in particular. We make the analogous assumptions for noncommutative geometries. The statement of ETH is that the expectation values of simple observables in high-energy eigenstates of chaotic systems are thermal. Furthermore, matrix elements of these observables behave as Gaussian random variables over energy bands. 

Our starting point is some classical configuration of matrices $X^1_{cl} \ldots X^D_{cl}$ that implicitly define $\mathcal{M}_{\theta}$ as in (as in \S\ref{ssec:rev-ncb}). We assume that the noncommutativity parameter is small, which translates to there existing a basis where the $X^i_{cl}$ are not very off-diagonal. More concretely, define $C$ as the weighted average Manhattan distance the elements of $X^i_{cl}$ from the main diagonal minimized over all choices of $U(N)$ bases. For the case of the fuzzy sphere $C=1$, as only the main diagonal and its two closest neighbors are occupied. The assumption of small noncommutativity parameter translates to the statement that $C \ll N$.

We define, as before,
\begin{equation}
    \delta X^i = X^i - X^i_{cl} = \sum_a \delta x_a Y_a^i, \quad K_{ij} := \langle \Pi^i \Pi^j \rangle = \sum_a \omega_a Y_a^i Y_a^j + O(g/N).
\end{equation}
If the kinetic term of the matrix lagrangian is given by
\begin{equation}
    L_K = \frac{1}{g^2}\Tr[\dot{X}^{i 2}] \leftrightarrow \frac{N}{g^2}\int \sqrt{G}d^d\sigma\, {\dot{a}^{i 2}},
\end{equation}
we note immediately that $\Pi^i$ (and therefore $K$) must scale as $g^{-2}$. Because the normal mode frequencies $\omega_a$ are proportional to $g^2$, $\langle \Tr[Q_{\Sigma}^2] \rangle$ will also scale as $g^{-2}$. For this reason the scaling of the entanglement with the coupling constant observed in \cite{Frenkel:2023aft} is generic.

Our goal now is to again argue as in the case of the fuzzy sphere that the noncommutative functions $K_{ij}$ are slowly varying, so we may again define an analogous operator $\mathcal{O}_Q$ that may be identified with a second order differential operator under the Moyal map. The functional form of $K_{ij}$ requires some assumptions about the normal modes of $Y_a^i$. For simplicity, we take the emergent noncommutative gauge theory on $\mathcal{M}_{\theta}$ to be free\footnote{Because high energy modes dominate the sum in $K_{ij}$, we only really need to assume an asymptotically free theory.}. This means we take the $Y_a^i$ to be eigenfunctions of the noncommutative laplacian:
\begin{equation}
[X_{cl}^i[X_{cl}^i,Y_a^j]] = \lambda_a Y_a^j \quad \forall a,j.
\end{equation}
We may therefore interpret the functions $Y^i_a$ as the wavefunctions of a spin-1 particle propagating on the surface of $\mathcal{M}_{\theta}$. We further assume the energy $\omega_a$ is a smooth function of $\lambda_a$ denoted $\omega(\lambda)$.

We separate $K_{ij}$ into slow and fast modes -
\begin{equation}\label{eqn:K-slow-fast}
K_{ij} = \sum_{\text{slow }a} \omega(\lambda_a) Y_a^i Y_a^j + \sum_{\text{fast } a} \omega(\lambda_a) Y_a^iY_a^j = K_{\text{slow},ij} + K_{\text{fast},ij}.
\end{equation}
$K_{\text{slow},ij}$ is manifestly a slowly varying function as it is entirely built out of low-energy modes. We now argue that to leading order in $N$ $K_{\text{fast},ij}$ is also slowly varying. We write it more explicitly as
\begin{equation}\label{eqn:K-sum}
(K_{fast,ij})_{nn'} = \sum_{\text{fast } a}\omega(\lambda_a) (Y_a^i)_{nn''}\overline{(Y_a^j)_{n'n''}}.
\end{equation}
It is here that we make use of the noncommutative analog of Berry's conjecture. We take the matrix elements of $Y_a^i$ for large $\lambda_a$ to be random complex variables with statistics that vary on the scale of breaking of translational invariance. This allows us treat the right hand side of \eqref{eqn:K-sum} as a random walk. 

On typical $\mathcal{M}_{\theta}$ we expect eigenvalue repulsion in the spectrum of the Laplacian, so there we do not have the large degenerate subspaces we enjoyed on the fuzzy sphere. However, we do expect that density of eigenvalues $\lambda$ to scale as $\lambda^{d-1}$. We therefore evaluate the sum in \eqref{eqn:K-sum} by first restricting to an energy band $\lambda_a \in [\lambda, \lambda + 1]$ and summing the bands. Over each band we take the objects $(Y_a^i)_{nn''}$ to be random variables with statistics
\begin{equation}
\text{EXP}_{\lambda}[(Y_a^i Y^j_a)_{nn'}] = (A_{ij}(\lambda))_{nn'}, \quad \text{VAR}_{\lambda}[(Y_a^i Y^j_a)_{nn'}] = O(1/N^2),
\end{equation}
where the $A_{ij}$ are the amplitudes of the slowly-varying envelopes. The constraint $\sum_i \Tr[Y_a^iY_a^i]=1$ ensures $A_{ij}$ are of size $O(1/N^2)$. The physical statement of this assumption is twofold:
\begin{enumerate}
\item Over the microcanonical ensemble, $Y^a_{n_1n'_1}$ and $Y^a_{n_2n'_2}$ are to good approximation independent if $n_2$ is far from $n_1$ or $n_2'$ is far from $n_1'$. We are working in a basis where the labels $n_i$ correspond to the approximate locations of $D$-branes, so we may rephrase this as saying bilocal degrees of freedom that don't share similar start and
end points are independent random variables.

Intuition for why we should expect this in MQM comes from the trace structure of the Lagrangian. Products of terms interacting in the Lagrangian must form closed index loops, so interaction between $\Pi_{n_1n_1'}^i$ and $\Pi^i_{n_2n_2'}$ must be mediated by degrees of freedom like $X^i_{n_1n_2}$ and $X^i_{n_1'n_2'}$ that stretch between $n_1$ and $n_2$ or $n_1'$ and $n_2'$ respectively. The mass of such degrees of freedom is proportional to the spatial separation of the regions of $\mathcal{M}_{\theta}$ represented by the $n_i$ indices, so will be suppressed to 0 when these indices are far apart. In eigenstates of chaotic systems, we expect observables that are weakly coupled to behave as independent random variables.

\item The expectation values $A_{nn'}$ vary on the scale of the breaking of translational invariance in the system as we vary the endpoints of the bilocal degrees of freedom. The \textit{raison d'\^{e}tre} of ETH is that in chaotic systems expectation values of observables in high-energy eigenstates are thermal. On Riemannian manifolds, we expect the thermal expectation values of local (or bilocal) observables -- such as the probability of a particle to occupy a coherent state centered on some position or momentum -- to vary no more quickly than the scale of curvature (see e.g. \cite{berry1983chaotic,srednicki1996thermal}).

\end{enumerate}

Putting everything together, the contribution to the sum will be given by summing a random walk:
\begin{equation}
\begin{split}
&K_{fast, ij} = \sum_{\lambda}\omega(\lambda) \left[D(\lambda)A_{ij}(\lambda) + O(1/\sqrt{D(\lambda)})\right] =\\
&=\left[\sum_{\lambda}\omega(\lambda)D(\lambda)A_{ij}(\lambda)\right] + O(1/N).
\end{split}
\end{equation}
The second equality is obtained by noting that the $O(1/\sqrt{D(\lambda)})$ corrections themselves form a random walk, so once we sum over $\lambda$ we have $O(N^2)$ steps in our random walk in total. We therefore again find that up to subleading in $C/N$ corrections, $K_{ij}$ represent functions $\kappa_{ij}(\vec{\sigma})$ that vary on the scale of curvature of $\mathcal{M}_{\theta}$, which we have taken to be much larger than the noncommutativity scale. 

What we have learned, therefore, is quite generally for emergent theories on $\mathcal{M}_{\theta}$ weakly coupled in the UV we may write
\begin{equation}
\begin{split}
&\left\langle\Tr[Q_{\Sigma}^2] \right\rangle = \Tr[K_{ij}\Theta_{\Sigma}[X^i_{cl},[X^j_{cl},\Theta_{\Sigma}]]] + O(1/N) = \Tr[\Theta_{\Sigma}\mathcal{O}_Q(\Theta_{\Sigma})],\\
&\mathcal{O}_{Q}(\cdot) = K_{ij}[X^i_{cl}[X^j_{cl},\cdot]] \leftrightarrow \tilde{G}^{ab}\tilde{\nabla}_a\tilde{\nabla}_b,
\end{split}
\end{equation}
for some metric structure $\tilde{G}$ on $\mathcal{M}_{\theta}$. For lengths $|\partial \Sigma_{(k)}|$ calculated with respect to $\tilde{G}$, the irrep dominating the state of edge modes again has the form shown in Fig. \ref{fig:fuzz-sphere-yt} for $U(1)$ emergent backgrounds and Fig. \ref{fig:fuzz-sphere-yt-2} for $U(q)$ emergent backgrounds. Unless forced by global symmetries of $\mathcal{M}_{\theta}$, $\mathcal{O}_Q$ will \textit{not} generically be related to the Laplacian, so the metric $\tilde{G}$ with respect to which we measure the length of entanglement cut is \textit{not} the same as the metric $G$ defined in \eqref{eqn:G-def}.

Although $\tilde{G}$ is not generally equal to $G$, this result is not vacuous - an arbitrary function from entangling surfaces to positive real numbers may not in general be interpreted as the lengths of these surfaces with respect to some metric. That we may find such a metric, and that it is related to $G$ by factors of smooth slowly-varying functions, is a sign of emergent local structure on the geometry despite the dominance of nonlocal effects in determining the functional form of $\kappa_{ij}$.

In the case where $\kappa_{ij}$ is proportional to the identity (in the $ij$ indices, not the $nn'$ indices, so $\kappa_{ij}(\vec{\sigma}) = \kappa(\vec{\sigma})\delta_{ij}$), we may interpret $\kappa(\vec{\sigma})$ as a local coupling constant setting $\tilde{G} = \kappa(\vec{\sigma})G$. This situation familiar from string theory, where the string frame metric setting the length and energy of the string is not the Einstein frame metric with respect to which we find area law entanglement behavior. More generally, we propose the interpretation of $K_{ij} \leftrightarrow \kappa_{ij}(\vec{\sigma})$ as a two-tensor of coupling constants that are dynamically determined by the generically nonlocal UV structure of the theory.

\section{Discussion}\label{sec:discussion}

We have defined the family of operators $\Tr[Q_{\Sigma}^{2p}]$ which probe the states of entanglement edge modes of MQM subsystems, with $Q_{\Sigma}$ defined as
\begin{equation}
2i\sum_{i=1}^D(\Pi^i_{\Sigma \bar{\Sigma}}X^i_{\bar{\Sigma}\Sigma} - X^i_{\Sigma \bar{\Sigma}}\Pi^i_{\bar{\Sigma}\Sigma}).
\end{equation}
We interpret $Q_{\Sigma}$ as the charge of the entanglement edge modes. That these modes are edge modes is to be expected, as in a system of interacting $D$-branes the $\Sigma\bar{\Sigma}$ and $\bar{\Sigma}\Sigma$ degrees of freedom are precisely string modes that mediate interaction between the two subregions. 

When the MQM wavefunction localizes to a noncommutative geometry $\mathcal{M}_{\theta}$, $Q_{\Sigma}^2$ is nicely determined as a noncommutative differential operator
\begin{equation}\label{eqn:disc-1}
\mathcal{O}_Q := K_{ij}[X^i[X^j,\cdot]], \quad K_{ij}:=\langle \Pi^i \Pi^j \rangle.
\end{equation}
We have argued that in general, the noncommutative functions $K_{ij}$ are smoothly varying on the scale of curvature of $\mathcal{M}_{\theta}$, implying that it tends toward a well behaved second order differential operator in the commutative limit. This differential operator is interpreted as the Laplacian of some metric structure $\tilde{G}$ on $\mathcal{M}_{\theta}$. We have argued this for emergent $U(1)$ and general $U(q)$ noncommutative gauge theories on $\mathcal{M}_{\theta}$.

This metric structure is not necessarily the same as the metric $G$ associated to the Laplacian $[X^i,[X^i,\cdot]]$ appearing in the kinetic term of the emergent field theory. The $K_{ij}$ we calculate are interpreted as emergent effective coupling constants, determined in general by the nonlocal UV/IR mixing structure of the high-energy normal modes of the theory. When $\mathcal{M}_{\theta}$ has sufficient global symmetry, $\tilde{G}$ \textit{is} constrained to match $G$. In such cases, we get honest boundary area-law behavior for the Young diagrams drawn in Figs. \ref{fig:fuzz-sphere-yt} and \ref{fig:fuzz-sphere-yt-2}.

\begin{figure}[h]
\centering
\begin{subfigure}{.4\textwidth}
  \centering
  \includegraphics[width=0.9\linewidth]{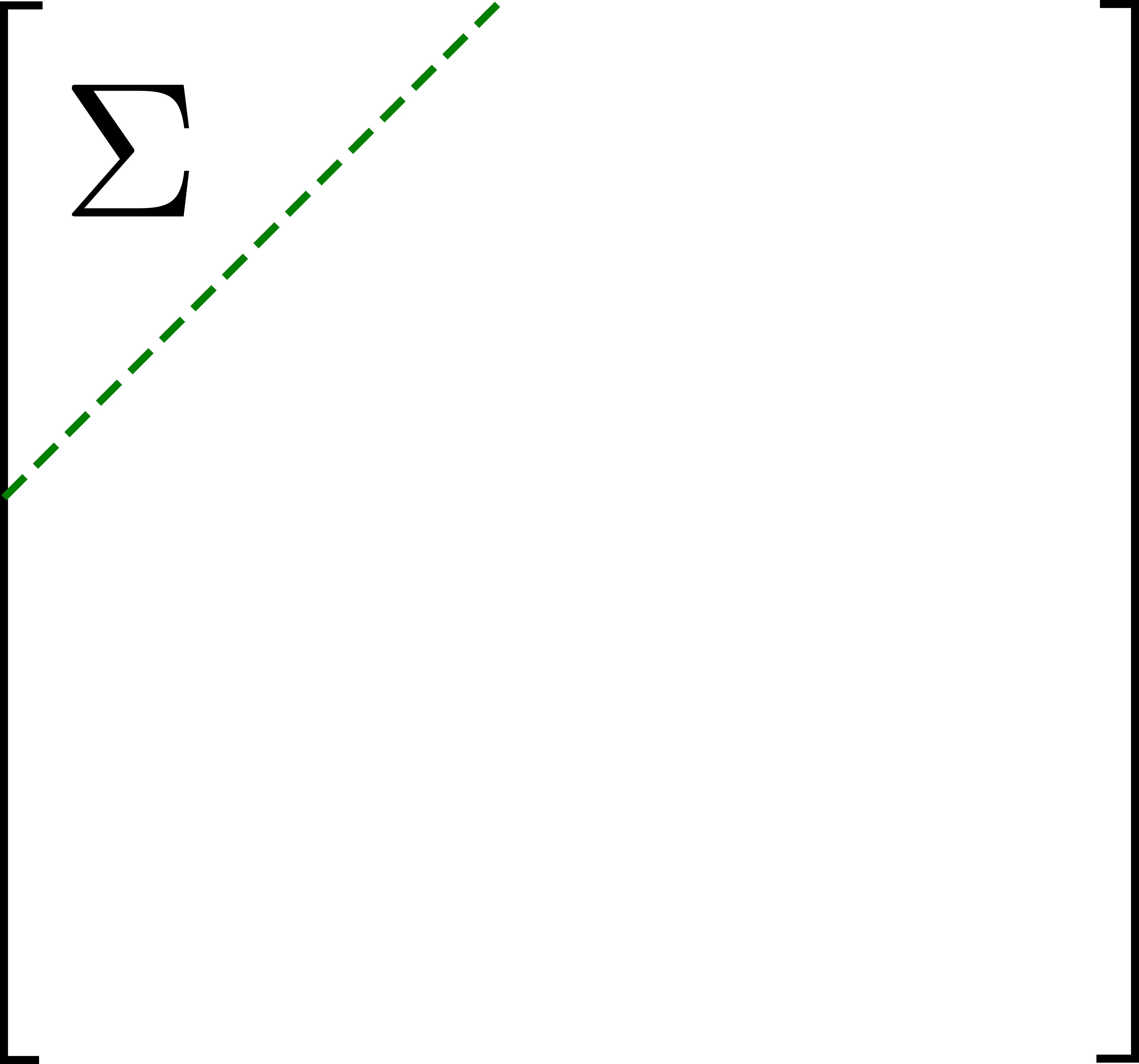}
  \caption{}
  \label{fig:oldcut}
\end{subfigure}%
\begin{subfigure}{.4\textwidth}
  \centering
  \includegraphics[width=0.9\linewidth]{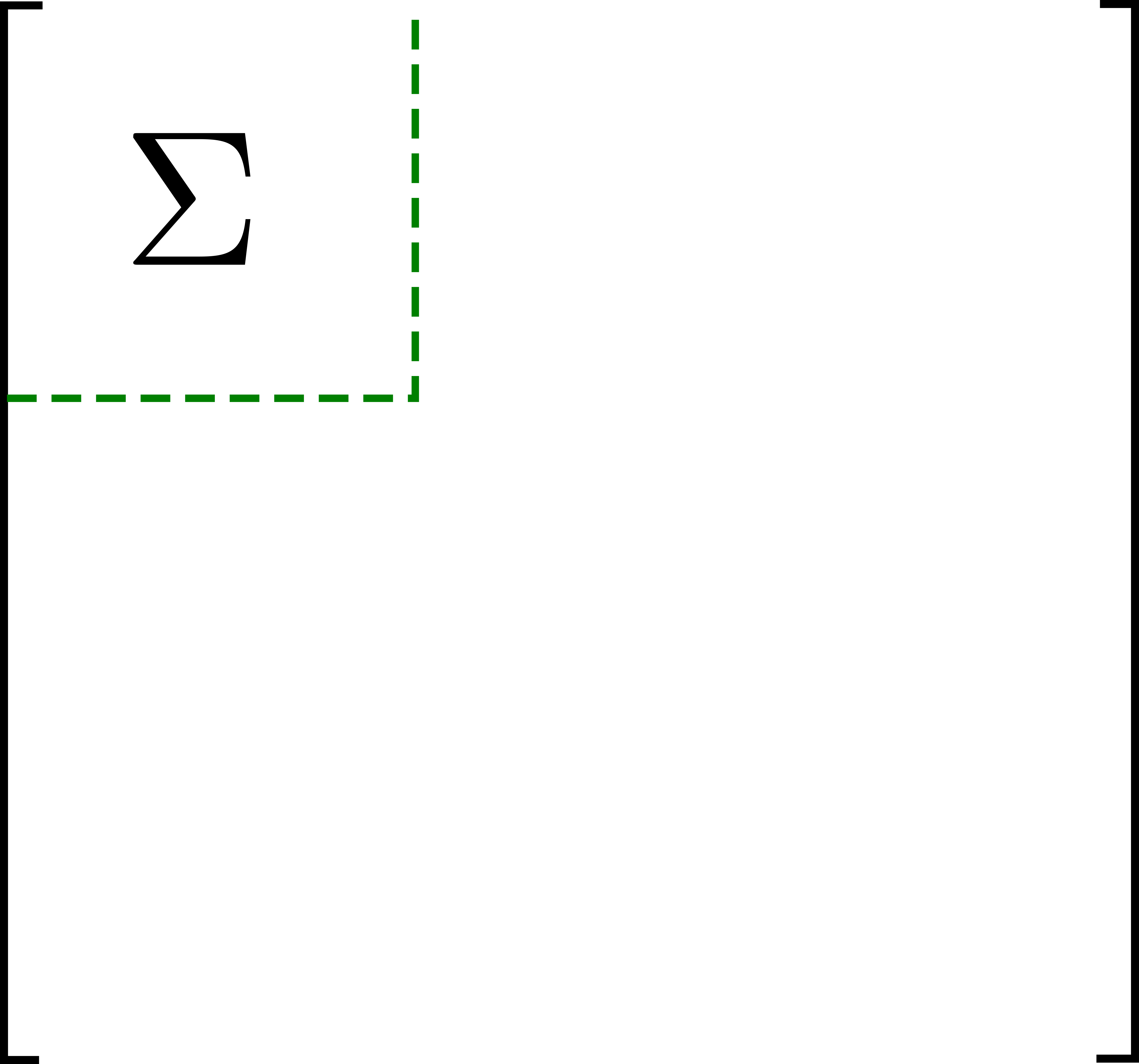}
  \caption{}
  \label{fig:newcut}
\end{subfigure}
\caption{Two choices of partitions of the matrix degrees of freedom for the cap subregion are pictured. On the left is the choice of partition considered in \cite{Karczmarek:2013jca}. On the right is the choice considered in \cite{Frenkel:2021yql,Frenkel:2023aft}, and this note. Both seem to give rise to area law behavior upon appropriate regularization, but the choice on the right makes the $U(M)$ symmetry of the subregion more manifest. It would be interesting to compare and contrast these two choices in greater detail for further work.}
\label{fig:two-cuts}
\end{figure}

It is interesting to observe that our choice of partition of the degrees of freedom is not the same as for previous computations of entanglement on the fuzzy sphere considered in \cite{Karczmarek:2013jca} (see Fig. \ref{fig:two-cuts}). The choice of a diagonal partition obscures the internal $U(M)$ symmetry of $\Sigma$ that is manifest in the block partition. It would be interesting to further compare and contrast these two choices of subalgebra.

For the case of the $U(q)$ emergent algebra, interesting further work would be to make more explicit contact with the edge mode algebra for commutative Yang-Mills theories found in \cite{Donnelly:2016auv}. In particular, one could compare to the chiral WZW edge mode algebra found for non-abelian matrix quantum hall states in \cite{dorey2016matrix,dorey2016matrix2}.

\subsection{Edge Modes in the Master Field}

In the `t Hooft limit, single trace operators factorize. This is enough to guarantee the existence of a master field \cite{Yaffe:1981vf,Gopakumar:1994iq} -- a collection of classical matrices $M^i$ that the wavefunction localizes to and that reproduce the expectation values of single trace observables up to $O(1/N)$ corrections. In particular, this suggests family of operators $\Tr[Q_{\Sigma}^{2p}]$ may have a nice bulk dual in holographic systems whose expectation value will be given by e.g.
\begin{equation}
\Tr[Q_{\Sigma}^{2p}] = \sum_a \omega_a \Tr[\left(M^i_{\Sigma \bar{\Sigma}}Y^i_{a,\bar{\Sigma}\Sigma} - Y^i_{a,\Sigma \bar{\Sigma}}M^i_{\bar{\Sigma}\Sigma}\right)^{2p}],
\end{equation}
where the normal modes $Y^i_a$ and their characteristic frequencies $\omega_a$ must be determined by the fully loop-corrected quadratic effective action around the classical minima $M^i$. 

A potential obstacle to a nice bulk dual is that $Q_{\Sigma}$ contains factors of the projection matrix $\Theta_{\Sigma}(\vec{X})$. The polynomial expansion of the Heaviside theta function contains arbitrarily large monomials in the $X^i$, and trace relations will expand $\Tr[Q_{\Sigma}^2]$ as a complex sum of multitrace operators. To combat this, one could perhaps smooth out the Heaviside theta functions by defining
\begin{equation}
\Tilde{\Theta}_{\Sigma}(X^i;A):= \frac{1}{2}\left(\tanh(A(f(X^i) - c)) + 1\right). 
\end{equation}

We work with these operators by taking $N$ large first, then taking $A$ large. It would be interesting to understand the bulk dual of $Q_{\Sigma}$, and whether the proposed connection to Susskind-Uglum string edge modes can be made manifest.

\section*{Acknowledgments}

I am partially supported by the NSF GRFP under grant no. DGE-165-6518. I would like to thank Amr Ahmadain, Jan Boruch, Sumit R. Das, Jackson Fliss, Joanna L. Karczmarek, Harold Steinacker, Stephen Shenker, Leonard Susskind, Sandip Trivedi, Cynthia Yan, Ming Yang, Zhenbin Yang, and especially Sean A. Hartnoll and Ronak M. Soni for many helpful and insightful discussions throughout the course of this work. I am additionally indebted to Dan S. Enceicu, Jackson Fliss, Sean A. Hartnoll, and Ronak M. Soni for detailed comments on an early draft.

\appendix
\section{Calculation Details for the Fuzzy Sphere}\label{app:fuzz-sph}

In this appendix, we explicitly show how to evaluate $\langle \Tr[Q_{\Sigma}^{2p}]\rangle$ for the $U(1)$ fuzzy sphere background.

A convenient basis for the algebra of functions on the fuzzy sphere is in terms of the matrix spherical harmonics $\hat{Y}_{jm}$. They are defined to satisfy
\begin{equation}
\begin{split}
[J^3,\hat{Y}_{jm}] &= m\hat{Y}_{jm}\\
\sum_i[J^i,[J^i,\hat{Y}_{jm}]] &= j(j+1)\hat{Y}_{jm}. 
\end{split}
\end{equation}
It is useful to note the properties
\begin{equation}\label{eqn:Y-hat-id}
\begin{split}
&j^{\pm}(j,m):=\sqrt{(j\mp m)(j \pm m + 1)}\\
&[J^{\pm},\hat{Y}_{jm}] = j^{\pm}(j,m),\\
&[J^{\pm},\hat{Y}_{jm}^{\dag}] = -j^{\pm}(j,m-1)\hat{Y}_{jm\mp 1}^{\dag}.
\end{split}
\end{equation}
Just like their commutative counterparts, the $\hat{Y}_{jm}$ furnish irreducible representations of the global $SO(3)$ symmetry of the fuzzy sphere, in the sense that
\begin{equation}
U(g)\hat{Y}_{jm}U^{\dag}(g) = D(g)^j_{mm'}\hat{Y}_{jm'}, 
\end{equation}
where $D^j_{mm'}(g)$ are the $SO(3)$ Wigner $D$ matrices. As shown in the appendix of \cite{Frenkel:2023aft}, the matrix elements $\hat{Y}_{jm}$ are precisely a discretization of the commutative spherical harmonic $y_{jm}(\theta,\phi)$ along the $m$th diagonal.

We choose a basis of solutions to \eqref{eqn:fuzzy-eom} where each normal mode $a$ has only one nonzero $y^3_{jm}$. We therefore drop the label $a$, as this label is implicit in which value of $y_{jm}^3$ is nonzero. The remaining coefficients are determined by the relation
\begin{equation}
    y^{\pm}_{j(m\pm 1)} = \pm \sqrt{\frac{(j\pm m + 1)(j \mp m)}{(\omega \pm m)^2}}y_{jm}^3.
\end{equation}
The modes that will dominate the ensuing sums will have $m,j-m \gg 1$, so we work with the approximation
\begin{equation}
\begin{split}
    &y^{\pm}_{j(m \pm 1)} \approx \pm y_{jm}^3\sqrt{\frac{j \mp m}{j \pm m}} \quad \omega = j + 1,\\
    & y^{\pm}_{j(m \mp 1)} \approx \pm y^3_{jm}\sqrt{\frac{j \pm m}{j \mp m}} \quad \omega = -j,\\
    &\sum_i |y_{a,jm}^i|^2 = \left(\frac{j+m}{j-m} + \frac{j-m}{j+m} + 1\right)|y^3_{jm}|^2 = |y_{jm}^3|^2\frac{3j^2 + m^2}{j^2-m^2} = 1.
\end{split}
\end{equation}
The last line fixes the normalization $y_{jm}^3$. We now apply the technology of sec. \ref{ssec:rev-ent}, focusing on regions defined by entanglement cuts that have radius of curvature large compared to the noncommutativity scale. More precisely, the expansion of the coordinate function defining our entanglement cut in terms of spherical harmonics satisfies
\begin{equation}\label{eq:f-exp}
\begin{split}
&f(\vec{x}) = \sum_{lm}c_{lm}y_{lm}(\vec{x}) \leftrightarrow F := \sum_{lm}c_{lm}\hat{Y}_{lm},\\
&\sum_{lm}|l||c_{lm}|^2 \ll N.
\end{split}
\end{equation}

This is the regime in which we expect geometric behavior in the entanglement, as the radius of curvature of the entanglement cut is much larger than the UV cutoff lengthscale. We take the coordinate matrix $F$ to have identical eigenvalues to $X^3_{cl}$, so we may write $F = U X^3_{cl} U^{\dag}$ for some unitary $U$. This is a convenient choice due to the constant spacing of eigenvalues of $X^3_{cl}$ - eigenfunctions of the differential operator $[X^i,\cdot]$ have a particularly nice form. In the large $N$ limit, conjugation under $U$ tends to the active area preserving diffeomorphism that takes the cap subregion studied in \cite{Frenkel:2023aft} to $\Sigma$.

The second Casimir of the edge modes of a subregion defined by $\Theta_{\Sigma}$ is given by \eqref{eqn:Q-sig-gen}
\begin{equation}\label{eq:qsig-sc}
\Tr[Q_{\Sigma}^2] \approx 4\nu^2 \sum_{ij}\sum_a \omega_a \Tr[J^i_{\Sigma \bar{\Sigma}}J^i_{\bar{\Sigma}\Sigma}Y^i_{a,\Sigma\bar{\Sigma}}Y^j_{a,\bar{\Sigma} \Sigma}].
\end{equation}
For low-curvature entanglement cuts of the sort described below \eqref{eq:f-exp}, we now show that to good approximation we may replace $\sum_a \omega_a Y^i_{a,\Sigma \bar{\Sigma}}Y^j_{a,\bar{\Sigma} \Sigma}$ inside the trace \eqref{eq:qsig-sc} with a sum over the unprojected normal modes $\sum_a \omega_a Y^i_{a}Y^j_{a}$. To see why this works, we first consider the cap subregion as a simple but illuminating example. We do this by considering functions on the sphere that have a definite angular momentum along the $\phi_F$ directions - i.e. matrices $\hat{Y}_{m_F}$ that satisfy
\begin{equation}
    [F,\hat{Y}_{m_F}] = m_F \hat{Y}_{m_F}, \quad \Tr[Y_{m_F}^{\dag}Y_{m'_F}] = \delta_{m_F,m'_F}.
\end{equation}
In the basis where $F$ is diagonal, these entries will be supported only on the $m_F$th diagonal. 

We first wish to consider the expansion of the coordinate functions $J^i$ in this basis
\begin{equation}
    J^i = \sum_{m_F} c^i_{m_F}\hat{Y}_{m_F}.
\end{equation}
It is useful to note that for any $\hat{Y}_{m_F}$ and $\hat{Y}_{m_F'}$, we have
\begin{equation}\label{eq:sig-drop}
    \Tr[\hat{Y}_{m_F\, \Sigma \bar{\Sigma}}\hat{Y}_{m_F\, \bar{\Sigma} \Sigma}\hat{Y}_{m_F'\, \Sigma \bar{\Sigma}}\hat{Y}_{m_F'\, \bar{\Sigma} \Sigma}] = \Tr[\hat{Y}_{m_F\, \Sigma \bar{\Sigma}}\hat{Y}_{m_F\, \bar{\Sigma} \Sigma}\hat{Y}_{m_F'}\hat{Y}_{m_F'}^{\dag}] \quad \text{iff } |m_F'| \geq |m_F|.
\end{equation}
This is proved directly by matrix multiplication. There is a large degeneracy amongst the eigenvalues $m_F$, but we leave the auxiliary summation index implicit. The only result that we need is that the $c^i_{m_F}$ have support primarily on small $m_F$ - specifically, on values of $m_F$ less than or equal to the curvature scale of the curvilinear coordinate $F$. We see this by considering the expressions
\begin{equation}
    \langle j_F^2 \rangle := \sum_{m_F} m_F^2 |c_{m_F}^i|^2 = \Tr[[F,J^i]^2].
\end{equation}
The latter expression is precisely the $F,F$ matrix element of the fuzzy sphere laplacian, which by the definition of a low-curvature entanglement cut is much smaller than $N^2$.

We now consider a similar expansion for a typical spherical harmonic in the $j$th representation. Denoting the curvilinear sphere coordinate perpendicular to $f$ as $\phi_f$, we write
\begin{equation}
    \langle m_F^2 \rangle = \Tr[[F,\hat{Y}_{jm}]^2] \leftrightarrow \int d^2 \sigma |\partial_{\phi_f} y_{jm}(\sigma)|^2.
\end{equation}
For a typical $\hat{Y}_{jm}$ with $j \gg m_F$, its derivatives in a random direction will have magnitude $O(j)$, so will in particular have $\langle m_F^2 \rangle = O(j^2)$. With all of this in mind, noting that the sum in \eqref{eq:qsig-sc} is dominated by $\hat{Y}_{jm}$ of order $N$ and using \eqref{eq:sig-drop}, we may write
\begin{equation}
    \Tr[Q_{\Sigma}^2] = 4\nu^2 \sum_{ij}\sum_a \omega_a \Tr[J^i_{\Sigma \bar{\Sigma}}J^i_{\bar{\Sigma}\Sigma}Y^i_{a}Y^j_{a}] + O(j_F/N).
\end{equation}
The $YY$ term in this expression is now independent of the entanglement cut we choose. In terms of matrix spherical harmonics, we define
\begin{equation}\label{eq:fij-def}
    \sum_a \omega_a Y_a^iY_a^j = \sum_a\sum_{j'm'}\sum_{jm}\bar{y}^i_{a,j'm'}y^j_{a,jm} \hat{Y}_{j'm'}^{\dag}\hat{Y}_{jm}=:K_{ij} \leftrightarrow \kappa_{ij}(\theta,\phi).
\end{equation}
$\kappa_{ij}(\theta,\phi)$ are a particular set of functions on $S^2$ entirely determined by the dynamics of the theory and independent of our choice of entanglement cut. They transform in the adjoint representation of the global $SO(3)$ in the sense that for $O \in SO(3)$ they satisfy
\begin{equation}\label{eq:K-trans}
    \kappa_{ij}(O \vec{x}) = O^l_i\kappa_{lm}(\vec{x}){O^{\dag}}_j^m.
\end{equation}

Despite the sum in \eqref{eq:fij-def} being dominated by high-energy modes, $\kappa_{ij}(\theta,\phi)$ is slowly varying in the sense that its derivatives $\partial_k \kappa_{ij}(\theta,\phi)$ are comparable to $\kappa_{ij}(\theta,\phi)$ itself. To prove this, consider a general sum of matrix spherical harmonics of the sort appearing in the definition of $K_{33}$:
\begin{equation}
\tilde{F}:=\sum_{m=-j}^j f(m)\hat{Y}_{jm}^{\dag}\hat{Y}_{jm}.
\end{equation}
We first observe $[J^3,\tilde{F}]=0$ as $\hat{Y}^{\dag}_{jm}\hat{Y}_{jm}$ is diagonal for all $m,j$. To check the other derivatives, we compute using \eqref{eqn:Y-hat-id}
\begin{equation}
\begin{split}
&[J^+,\tilde{F}] = \sum_{m=-j}^jf(m)\left([J^+,\hat{Y}^{\dag}_{jm}]\hat{Y}_{jm} + \hat{Y}_{jm}^{\dag}[J^+,\hat{Y}_{jm}]\right)=\\
&=\sum_{j=-m}^mf(m) \left(-j^+(j,m-1)\hat{Y}^{\dag}_{jm-1}\hat{Y}_{jm} + j^{+}(j,m)\hat{Y}_{jm}^{\dag}\hat{Y}_{jm+1}\right).
\end{split}
\end{equation}
We now shift the index of the second term in the sum to write
\begin{equation}
\begin{split}
&[J^+,\tilde{F}]=\sum_{m=-j}^{j-1} \left(f(m)-f(m+1)\right)j^+(j,m)\hat{Y}^{\dag}_{jm}\hat{Y}_{jm+1} \approx\\
&\approx\sum_{m=-j}^{j-1}f'(m)j^+(j,m)\hat{Y}^{\dag}_{jm}\hat{Y}_{jm+1} + O(f''(m)/f'(m)).
\end{split}
\end{equation}

\begin{figure}[h]
\begin{center}
\includegraphics[width=0.6\textwidth]{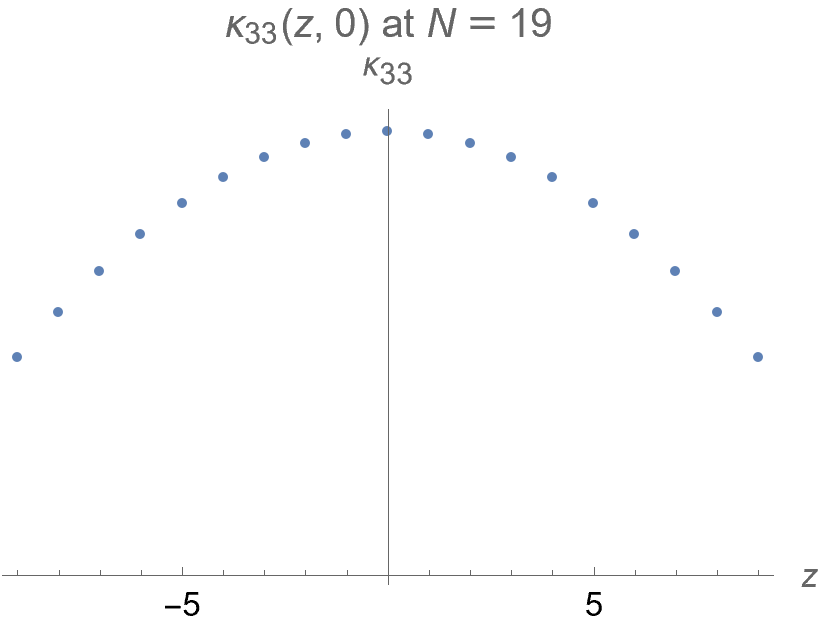}
\caption{$\kappa_{33}(z,0)$, evaluated numerically using \eqref{eq:fij-def} for $N=19$. One can observe that despite the sum being dominated by high energy modes and the UV/IR mixing issues observed in \cite{Karczmarek:2013jca,Frenkel:2023aft}, this is a slowly varying function on the fuzzy sphere.}
\label{fig:k33-plot}
\end{center}
\end{figure}

For $K_{33}$, we have
\begin{equation}
f(m)=j\frac{1-(m/j)^2}{3+(m/j)^2} \implies \frac{f'(m)}{f(m)} = O(1/j), \quad j^+(j,m)=O(j).
\end{equation}
The $O(1/j)$ scaling of $f(m)$ cancels the $O(j)$ scaling of the spherical harmonic derivatives, and therefore the derivatives of $K_{33}$ are comparable to $K_{33}$ itself. In this sense, $K_{33}\leftrightarrow \kappa_{33}(\theta,\phi)$ is a slowly varying function on the fuzzy sphere. To check our argument, we also numerically evaluate $K_{33}$ and plot its eigenvalues for $N=19$ in Fig. \ref{fig:k33-plot}. Similar behavior may be shown for the remaining $K_{ij}$ using the same methods or via \eqref{eq:K-trans}. 

With all of this in mind, we may now write
\begin{equation}\label{eqn:int-by-parts}
\begin{split}
    &\Tr[Q_{\Sigma}^2] = 2\nu^2 \sum_{ij}\Tr[[\Theta_{\Sigma},J^i][\Theta_{\Sigma},J^j]K_{ij}]=\\
    &= 2\nu^2\sum_{ij}\Tr[\Theta_{\Sigma}\left([J^i,[J^j,\Theta_{\Sigma}]]K_{ij} + [J^i,\Theta_{\Sigma}][J^j,K_{ij}]\right)].
\end{split}
\end{equation}
Precisely because $\kappa_{ij}$ has $O(1)$ derivatives, we may drop the second term that has noncommutative derivatives acting on $K_{ij}$ instead of the delta function piece. We therefore arrive at the expression
\begin{equation}
    \Tr[Q_{\Sigma}^2] \approx 2\nu^2\sum_{ij}\Tr[\Theta_{\Sigma}\left([J^i,[J^j,\Theta_{\Sigma}]]K_{ij}\right)] \leftrightarrow 4\nu^2 \int d^2\sigma \chi_{\Sigma} \kappa_{ij}\tilde{L}_i \tilde{L}_j \chi_{\Sigma}.
\end{equation}

We have now arrived at \eqref{eqn:OQ-def}.

\bibliographystyle{JHEP}
\bibliography{refs}
 
\end{document}